\documentclass[12pt]{iopart}

\usepackage{graphicx}
\usepackage[all]{xy}
\usepackage{epstopdf}
\usepackage{pdflscape}			
\usepackage{cite}				

\usepackage{color}

\renewcommand{\eref}[1]{equation (\ref{#1})}

\begin{document}

\title{Phase Diagrams of Systems of $2$ and $3$ levels in the presence of a Radiation Field \\}

\author{Eduardo
Nahmad-Achar, Sergio Cordero, Octavio Casta\~nos and Ram\'on L\'opez-Pe\~na}

\address{Instituto de Ciencias Nucleares, Universidad Nacional
Aut\'onoma de M\'exico, Apdo. Postal 70-543 M\'exico 04510 D.F.}

\ead{nahmad@nucleares.unam.mx}


\begin{abstract}

	We study the structure of the phase diagram for systems consisting of $2$- and $3$-level particles dipolarly interacting with a $1$-mode electromagnetic field, inside a cavity, paying particular attention to the case of a {\it finite} number of particles, and showing that the divergences that appear in other treatments are a consequence of the mathematical approximations employed and can be avoided by studying the system in an exact manner quantum-mechanically or via a catastrophe formalism with variational trial states that satisfy the symmetries of the appropriate Hamiltonians.

These variational states give an excellent approximation not only to the exact quantum phase space, but also to the energy spectrum and the expectation values of the atomic and field operators. Furthermore, they allow for {\it analytic} expressions in many of the cases studied. We find the loci of the transitions in phase space from one phase to the other, and the order of the quantum phase transitions are determined explicitly for each of the configurations, with and without detuning. 

We also derive the critical exponents for the various systems, and the phase structure at the {\it triple point} present in the $\Xi$-configuration of $3$-level systems is studied.

\end{abstract}


\maketitle


\section{Introduction}

While some observed phenomena such as the Rabi cycles in $2$-state quantum systems may be explained by a semi-classical theory, other occurrences such as the revival of the atomic population inversion after its collapse~\cite{cummings65, eberly80, rempe87} are quantum effects derived as a consequence of the discreteness of the field states. The revival property appears as well, for instance, in the dynamics of electron currents in monolayer graphene subject to a magnetic field~\cite{romera09}. Even fractional revivals have been identified with information entropies in different physical systems of interest~\cite{romera07}. (A review of the formalism required to understand some aspects of the revival behaviour is presented in~\cite{robinett04}.)  These and other purely quantum effects need to be studied through a quantum optics model such as the Jaynes-Cummings model (JCM)~\cite{jaynes63}, which describes the behaviour of a $2$-level system in the presence of a quantised radiation field. This model works very well when the radiation field and the system energy gap are close to one another and of the order of optical frequencies ($\sim 10^{15}$ Hz); this approximation is the so-called {\it rotating wave approximation} (RWA). The extension of the model to many ``atoms'' or systems is the Tavis-Cummings model (TCM)~\cite{tavis69}, and the removal of the RWA approximation including the so-called {\it counter-rotating} terms leads to the Dicke model (DM)~\cite{dicke54}, which describes the interaction of a single mode quantized radiation field with a sample of $N_A$ two-level atoms located inside an optical cavity, in the dipolar approximation (i.e., located within a distance smaller than the wavelength of the radiation). (Hereafter we will refer to ``atoms'', but the theory applies to any finite-level system, including spin systems and Bose-Einstein condensates.) The Dicke Hamiltonian has the expression
\begin{equation}
	\fl\qquad
H = \hbar \omega_F\, a^\dagger a + \tilde\omega_A\, J_z + \frac{\tilde\gamma}{\sqrt{N_A}} (a^\dagger J_- + a J_+) + \frac{\tilde\gamma}{\sqrt{N_A}} (a^\dagger J_+ + a J_-).
\label{DickeHamiltonian}
\end{equation}
Here, $N_A$ is the number of particles; the first term in the rhs represents the field Hamiltonian, where $\omega_F$ is the field frequency and $a^\dagger,\,a$ are the creation and annihilation photon operators; the second term represents the atomic Hamiltonian, with $\tilde\omega_{A}$ the atomic energy-level difference, and $J_{z}$ the atomic relative population operator. The $2$ last terms represent the interaction Hamiltonian; we have written them separately in order to differentiate the {\it rotating} term (first), with $J_{\pm}$ the atomic transition operators, and the {\it counter-rotating} term (second). $\tilde\gamma$ is the dipolar coupling constant.

The parameters appearing in (\ref{DickeHamiltonian}) are related to the physical properties of the atom/system, and have dimensions as shown in Table~\ref{t1}, where $d$ is the dipole moment of the atom, $e$ the electron charge, and $\rho$ the atomic density inside the quantisation volume. It is convenient to redefine $\omega_A =\frac{\tilde\omega_A}{\omega_F}, \ \gamma =\frac{\tilde\gamma}{\omega_F}$, and take $\omega_F=1$ (i.e., measure frequency in units of the field frequency), which we do hereafter.

\begin{table}
	\caption{Hamiltonian parameters in terms of physical parameters of the system.}
	\vspace{0.1in}
	\begin{center}
\begin{tabular}{l|l}
	dimensional & dimensionless \\[0.5ex]
	\hline
	\rule[-5mm]{0mm}{15mm}
	$\displaystyle\tilde\gamma = \tilde\omega_A \,d \,\sqrt{\frac{2\pi\rho}{\hbar\,\omega_F}}\ \ \hbox{[freq]}$ & $\displaystyle\gamma = \omega_A\,d\,\sqrt{\frac{2\pi\rho}{\hbar\,\omega_F}}$
\\[5mm]
	$\displaystyle\tilde\kappa\,\tilde\gamma^2 = \frac{e^2}{2m}\,\frac{2\pi\rho}{\omega_F N_A}\ \ \hbox{[freq]}$ & $\displaystyle\kappa\,\gamma^2 = \frac{\pi\,e^2\,\rho}{m\,\omega_F^2 N_A}$ 
\\[5mm]
	$\displaystyle\tilde\kappa = \frac{e^2\,\hbar}{2m\,d^2\tilde\omega_A^2 N_A}\ \ \hbox{[1/freq]}$ & $\displaystyle\kappa = \frac{e^2\,\hbar}{2m\,d^2\tilde\omega_A\omega_F\,N_A}$
\end{tabular}
	\end{center}
\label{t1}
\end{table}

The expression~(\ref{DickeHamiltonian}) was derived from a multipolar expansion of the dipole interaction with the electromagnetic field. A different derivation related to the radiation gauge, where the long wavelength approximation is considered as well as the approximation to $2$-level systems, leads to an extra {\it diamagnetic} term quadratic in the electromagnetic vector potential $A$, of the form $\frac{\tilde\kappa \tilde\gamma^2}{N_A} (a^\dagger + a)^2$, with $\tilde\kappa \tilde\gamma^2$ the {\it diamagnetic coupling constant}. This has led to some confusion in the literature as to the correct expression to use. Both the multipolar and the radiation Hamiltonians are related by a unitary gauge transformation, thus yielding the same physics; it is the approximation to $2$-level systems that breaks this symmetry (cf.~\cite{camop2} for details). When using the Hamiltonian derived from the radiation gauge, the Thomas-Reiche-Kuhn sum rule would place contradictory bounds to the parameters of the model~\cite{rzazewski75}; furthermore, since the coupling strength $\tilde{\gamma}$ is much smaller than the atomic level separation $\tilde{\omega}_A$ for optical systems, it was believed that gauge invariance {\it requires} the presence of the diamagnetic term~\cite{knight78}. To the benefit of the Hamiltonian in~(\ref{DickeHamiltonian}), not only has a very strong case been made in its favour as a consistent description of the interaction of a one-mode light field with the internal excitation of atoms inside a cavity~\cite{domokos12}, but experimental results indicate that transitions apparently forbidden by the {\it no-go theorem} from the sum-rule mentioned above can actually be observed~\cite{baumann10, nagy10} by using Raman transitions between ground states in an atomic ensemble.

\begin{figure}
		\begin{center}
		\scalebox{0.6}{\includegraphics{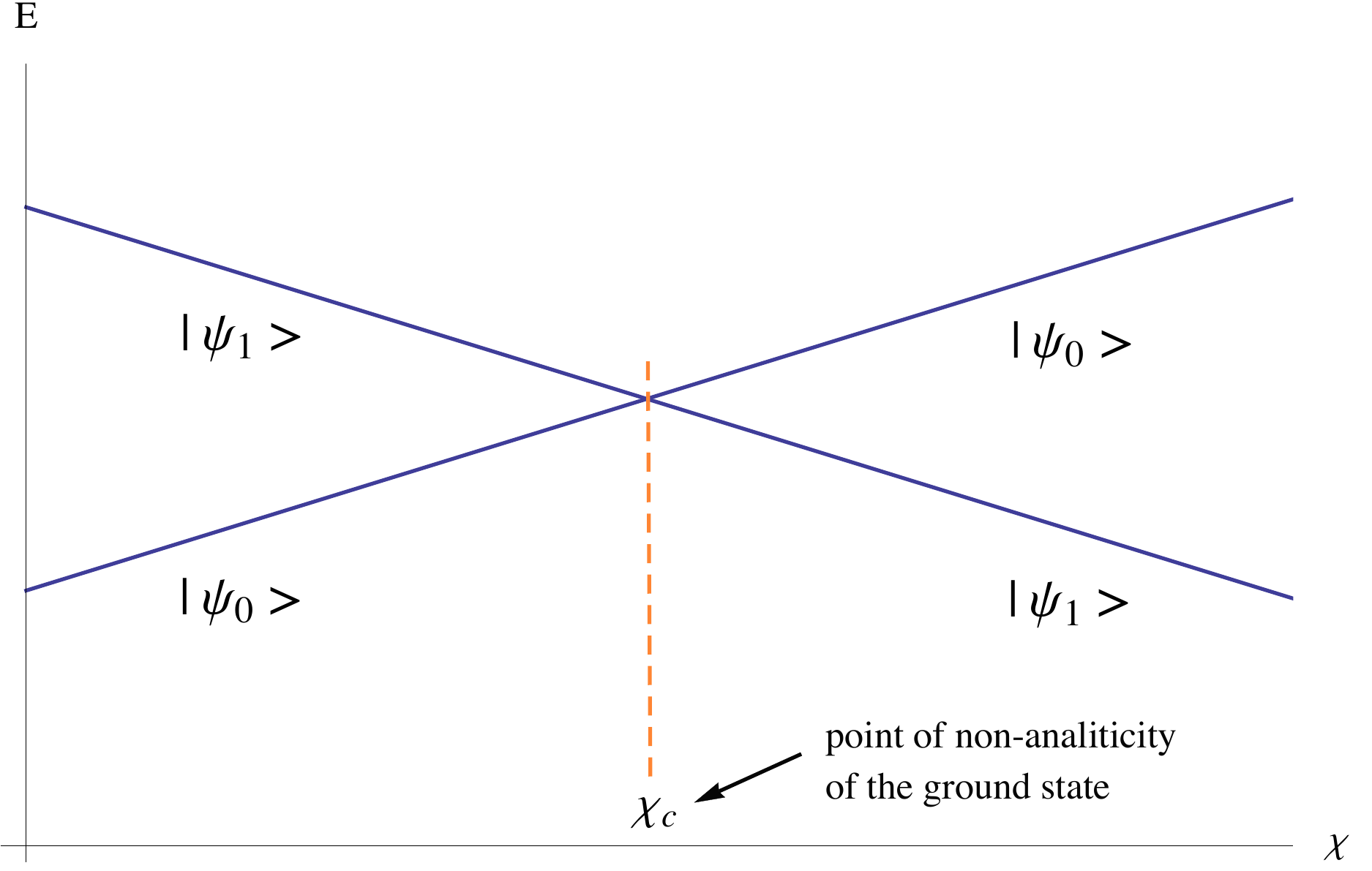}}
		\end{center}
		\caption{Energy-level crossing at critical point $\chi_c$ of non-analiticity of the ground state.}
	\label{Ecrossing}
	\end{figure}

An important feature of atom-field interactions is the presence of phase transitions~\cite{hepp73} from the {\it normal} to a {\it collective} behaviour: effect involving all $N_A$ atoms in the sample, where the decay rate is proportional to $N_A^2$ instead of $N_A$ (the expected result for independent atom emission). Quantum fluctuations may drive a change in the ground state of a system, even at zero temperature, $T=0$. A simple way to see this is to consider a Hamiltonian $H(\chi)$, whose degrees of freedom vary as a function of a dimensionless coupling parameter $\chi$. The ground-state energy of $H(\chi)$ would generally be a smooth, analytic function of $\chi$~\cite{sachdev}. Exceptions occur, for example, in the case when $\chi$ couples only to a conserved quantity
\begin{equation}
H(\chi) = H_0 + \chi\,H_1
\end{equation}
where $H_0$ and $H_1$ commute. Then $H_0$ and $H_1$ can be simultaneously diagonalised, but while the eigenvalues vary with $\chi$, the eigenfunctions are independent of $\chi$. We can then have a level-crossing where an excited 
level becomes the ground state at a certain {\it critical} value of the coupling $\chi=\chi_c$ (cf. Fig.~\ref{Ecrossing}).

For $\chi \ll \chi_c$ and $\chi \gg \chi_c$ the ground state of the system is clear; for $\chi \rightarrow \chi_c$ terms start to compete, and the system would undergo a phase transition: we say that the limiting states {\it realise} distinct quantum phases. The crossing of levels in the spectrum of a physical system is an indication of a first order transition while, in general, the second order ones correspond to other causes (e.g., avoided crossings) and they are continuous. Each phase is a region of analyticity of the free energy per particle, and different phases are separated by {\it separatrices} which are singular loci of the free energy. Thus, the study of the phase diagram of a system is an important means to understand its behaviour.

There have been various contributions to the study of phase transitions in $2$-level systems~\cite{reslen05, zanardi06, goto08, emary03}. In particular, the Husimi function has been used for phase space analysis~\cite{romera12a} and entropic uncertainty relations to detect quantum phase transitions~\cite{romera12b}. Here, we want to stress the role of the catastrophe formalism to determine significant changes in the ground state of the system under small changes in the parameters of the model. Quantum phase transitions and stability properties have been extensively studied through the catastrophe formalism and the coherent states theory~\cite{gilmore81, scrip, scrip2, papercorto, paperextenso, universal, camop1, crossovers}. In particular, as these quantum systems cannot be solved analytically in an exact manner (except in the thermodynamic limit), in the latter references a procedure based on the use of the fidelity susceptibility of neighbouring states was established to determine with fine precision the location of the separatrices (but see also~\cite{zanardi06, gu10}).

For applications such as quantum memories and other quantum information and quantum optics purposes it seems more appropriate to use $3$-level atoms. Furthermore, approximations to $3$-level systems in the $\Lambda$ configuration are plentiful: e.g., alkali metals, as confirmed by the electromagnetically induced transparency effect. For practical applications, off-resonant systems protect one from spontaneous emission and have thus been favored because of their advantage when subjected to coherent manipulations; in fact, schemes have been presented for various quantum gates using 3-level atoms and trapped ions~\cite{yi03, jane02}. The study of $3$-level systems thus deserves attention. In particular, the importance of their phase diagrams has drawn the attention of some authors~\cite{cordero1, baksic}. In~\cite{cordero1} the energy surface method was applied to obtain an estimation of the ground state energy and the phase diagrams as well as the order of the phase transitions, in the three configurations, using the multipolar Hamiltonian. The results were compared with those of the exact quantum solution. In~\cite{baksic} the radiation Hamiltonian containing the diamagnetic term is used, in the Holstein-Primakoff realisation. They analyse the phase diagram of the three configurations in the thermodynamic limit, taking care of the regions where the Thomas-Reiche-Kuhn (TRK) sum rule holds, and they show that transitions from the normal to the collective regimes are possible even when the TRK rule is satisfied; this is in direct contrast to the situation of $2$-level systems.

\begin{figure}
		\begin{center}
		\scalebox{1.4}{\includegraphics{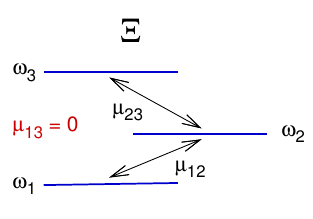}} \quad
		\scalebox{1.4}{\includegraphics{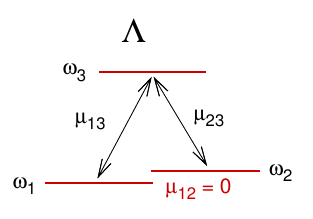}} \quad
		\scalebox{1.4}{\includegraphics{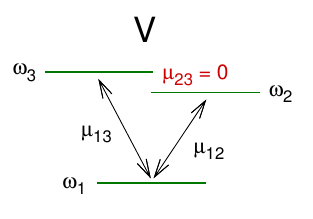}}
		\end{center}
		\caption{Atomic configurations $\Xi$, $\Lambda$ and $V$. The $i$-th level frequency is denoted by $\omega_i$ with the convention $\omega_1\leq\omega_2\leq\omega_3$, and the coupling parameter between levels $i$ and $j$ is $\mu_{ij}$.}
	\label{configurations}
	\end{figure}

Here, for the aforementioned reasons, we will consider the multipolar Hamiltonian and we will make use of the catastrophe formalism to study $3$-level systems. Their Hamiltonian may be written as~\cite{LAOP}
\begin{equation}
	H = H_D + H_{int},
	\label{Hamiltonian3Levels}
\end{equation}
where $H_D$ and $H_{int}$ are the diagonal and interaction contributions, respectively given by
\begin{eqnarray}
	\fl
	H_D &=& \Omega \,a^\dag\,a + \omega_1\, A_{11} + \omega_2\, A_{22} + \omega_3\, A_{33},\label{HD} \\[3mm]
	\fl 
	H_{int} &=& -\frac{1}{\sqrt{N_A}} \left[ \mu_{12}\left(A_{12} + A_{21}\right) + \mu_{13}\left(A_{13} + A_{31}\right) + \mu_{23}\left(A_{23} + A_{32}\right) \right]\, \left(a^\dag + a \right).
	\label{Hint}
\end{eqnarray}
Here $a^\dag,\,a$ are as before the creation and annihilation electromagnetic field operators, and $A_{ij} = \sum_{s=1}^{N_A}\,A_{ij}^{(s)}$ the collective matter operators obeying the $U(3)$ algebra
\begin{equation}
\left[A_{ij},\,A_{lm}\right] = \delta_{jl}\,A_{im}-\delta_{im}\,A_{lj},
\end{equation}
with a possible realisation $A_{ij}^{(s)} = \vert i^{(s)} \rangle\,\langle j^{(s)} \vert$, and the total number of atoms is given by
\begin{equation}
N_A=\sum_{k=1}^3 A_{kk}.
\end{equation}
The $i$-th level frequency is denoted by $\omega_i$ with the convention $\omega_1\leq\omega_2\leq\omega_3$, and the coupling parameter between levels $i$ and $j$ is $\mu_{ij}$. The different atomic configurations are chosen by taking the appropriate value $\mu_{ij}=0$ (cf. Fig.~\ref{configurations}).

We have written $\Omega$ (instead of $\omega_F$) for the frequency of the radiation field. The way in which equations (\ref{HD},\ref{Hint}) are written lends itself to be easily generalised for a system of $n$-level atoms interacting with $m$-modes of a radiation field
\begin{equation}
	\fl
	\qquad H_{\hbox{\footnotesize{general}}} = \sum_{\ell=1}^n \Omega_\ell\, a^\dag_\ell\, a_\ell + \sum_{k=1}^n \omega_k\, A_{kk} - \sum_{j<k}\, \sum_{\ell=1}^m\, \frac{1}{\sqrt{N_A}}\, \mu_{jk}^\ell\, \left( A_{jk} + A_{kj} \right)\, \left( a^\dag_\ell + a_\ell \right)
	\label{Hamiltonian_nLevels_mmodes}
\end{equation}
where the values of $j,\,k,\,\ell$ are determined by the possible transitions according to the specific atomic configuration, and where we have $N_A=\sum_{k=1}^n A_{kk}.$

In this work we shall review and extend the study of the phase diagrams for $2$- and $3$-level quantum systems consisting of a {\it finite} number of atoms interacting through a $1$-mode electromagnetic field. We show how the use of variational trial states that are adapted to the symmetry of the system Hamiltonian give an excellent approximation not only to the exact quantum phase space, but also to the energy spectrum and the expectation values of the atomic and field operators. When in the RWA approximation, the total number of excitations is an integral of motion of the system; using trial states adapted to the symmetry of the Hamiltonian then means essentially {\it projecting} onto this integral of motion. In the full model (rotating and counter-rotating terms), however, it is the {\it parity} in the number of excitations that is conserved, and to obtain {\it symmetry-adapted} states (SAS) we therefore take linear combinations of coherent states of the same parity.

These symmetry adapted states were first used in~\cite{manko74}, named ``even and odd coherent states'', as nonclassical states for the study of singular non-stationary quantum-mechanical harmonic oscillators, and later to discuss the properties of the tomographic representation of quantum mechanics~\cite{manko01, manko03}. Here, we use them to look in detail at the structure of the phase diagram and the behaviour of the phase changes. We also present some virtues and limitations of these symmetry-adapted states, use the fidelity and the fidelity susceptibility of neighbouring quantum states to find the loci of the transitions in phase space from one phase to the other, and derive the critical exponents for the various systems.
\vspace{0.1in}

{\it This work is dedicated with great appreciation to Professors Vladimir and Margarita Man'ko in their joint $150$-year celebration, for their numerous contributions to the development and promotion of quantum optics and mathematical physics.}

\section{Two-level Systems}

The simplest completely soluble quantum-mechanical model of one $2$-level atom in an electromagnetic field is described by the Jaynes-Cummings (JCM) model~\cite{jaynes63}. This, and its generalisation to $N_A$ identical $2$-level atoms given by Tavis and Cummings~\cite{tavis69}, the TCM model, were fundamental to study basic properties of quantum electrodynamics and to understand phenomena like the existence of collapse and revivals in the Rabi oscillations (observed experimentally for the first time in 1987~\cite{rempe87}). Both the JCM and the TCM models discard the terms in the Hamiltonian which do not conserve the total number of excitations of the field plus matter by using the RWA approximation. When these terms are considered we obtain the full Dicke model (DM)~\cite{dicke54}. In this Section we consider the phase diagrams presented by these models for the ground state, both in the case of a finite number of atoms $N_A$ and in the thermodynamic limit, by making use of the catastrophe formalism to determine when significant changes to the ground state occur for small changes of the external environment (the parameters of the model). The influence of the phase transitions on the behaviour of observables of interest for the matter and the field are also presented.

The choice of the use of the catastrophe formalism allows us to obtain analytic descriptions for the phase diagram in parameter space, which distinguishes the normal and collective regions, and which gives us all the quantum phase transitions of the ground state from one region to the other as we vary the interaction parameters (the matter-field coupling constants) of the model, in functional form. This approach thus allows also for the study of the asymptotic behaviour in any of the quantities of interest: the number of particles, the constants of motion, and the interaction parameters themselves.

Catastrophe theory derives from the research of Ren\'e Thom in topology and differential analysis on the structural stability of differentiable maps~\cite{thom}. Dissipative systems, for example, always reach equilibrium; this equilibrium is characterised by a certain function $\mu(x)$ which at $x$ represents the minimum of usually the {\it energy} of the system, and when this minimum $\mu(x)$ is stable $x$ will be a regular point in the space of parameters describing the system. But when the energy changes abruptly at $\mu(x)$ due to slight disturbances the local minimum is destroyed in a neighbourhood of $x$, $\mu(x)$ ceases to be an attractor of the dynamics, and $x$ is a {\it catastrophic} point: the state of the system will present sudden jumps from $x$ to another point $x'$ (another attractor) and back. The dynamics of the system thus bifurcates. It is these bifurcations that we are interested in studying analytically.

\subsection{The Jaynes-Cummings and the Tavis-Cummings Models}

A $2$-level system of $N_A$ atoms interacting dipolarly with an electromagnetic field of frequency $\omega_F$ is described by the Tavis-Cummings Hamiltonian~\cite{tavis69}, which we may write as
\begin{equation}
	H = \frac{1}{N_A}\omega_F\,a^\dagger a + \frac{\omega_A}{N_A} J_z + \frac{\gamma}{\sqrt{N_A} N_A} (a^\dagger J_- + a J_+)
	\label{TCHamiltonian}
\end{equation}
where we have set $\hbar=1$ and all quantities are dimensionless. We have also divided the expression by $N_A$ in order to consider an intrinsic Hamiltonian, which we do hereafter. We can set $\omega_F=1$ (i.e., measure frequency in units of the field frequency), and define a {\it detuning parameter} $\Delta = \omega_F - \omega_A = 1 - \omega_A$; thus, $\Delta = 0$ when particles and field are in resonance and $\Delta \neq 0$ when away from resonance. It is convenient to introduce $\hat\Lambda = \sqrt{\hat{J}^2 + 1/4} -1/2 + \hat{J}_z + a^{\dagger}a$ because it turns out to be an integral of motion for the system. Its eigenvalues are $\lambda = \nu + m + j$, with $j=N/2$, $j+m$ the number of atoms in their excited state, and $\nu$ the number of photons. The Hamiltonian can then be rewritten as
\begin{equation}
H = \frac{1}{N_A}\Lambda - \frac{\Delta}{N_A} J_z + \frac{\gamma}{\sqrt{N_A} N_A} (a^\dagger J_- + a J_+).
\end{equation}

\begin{figure}
	\begin{center}
	\scalebox{0.45}{\includegraphics{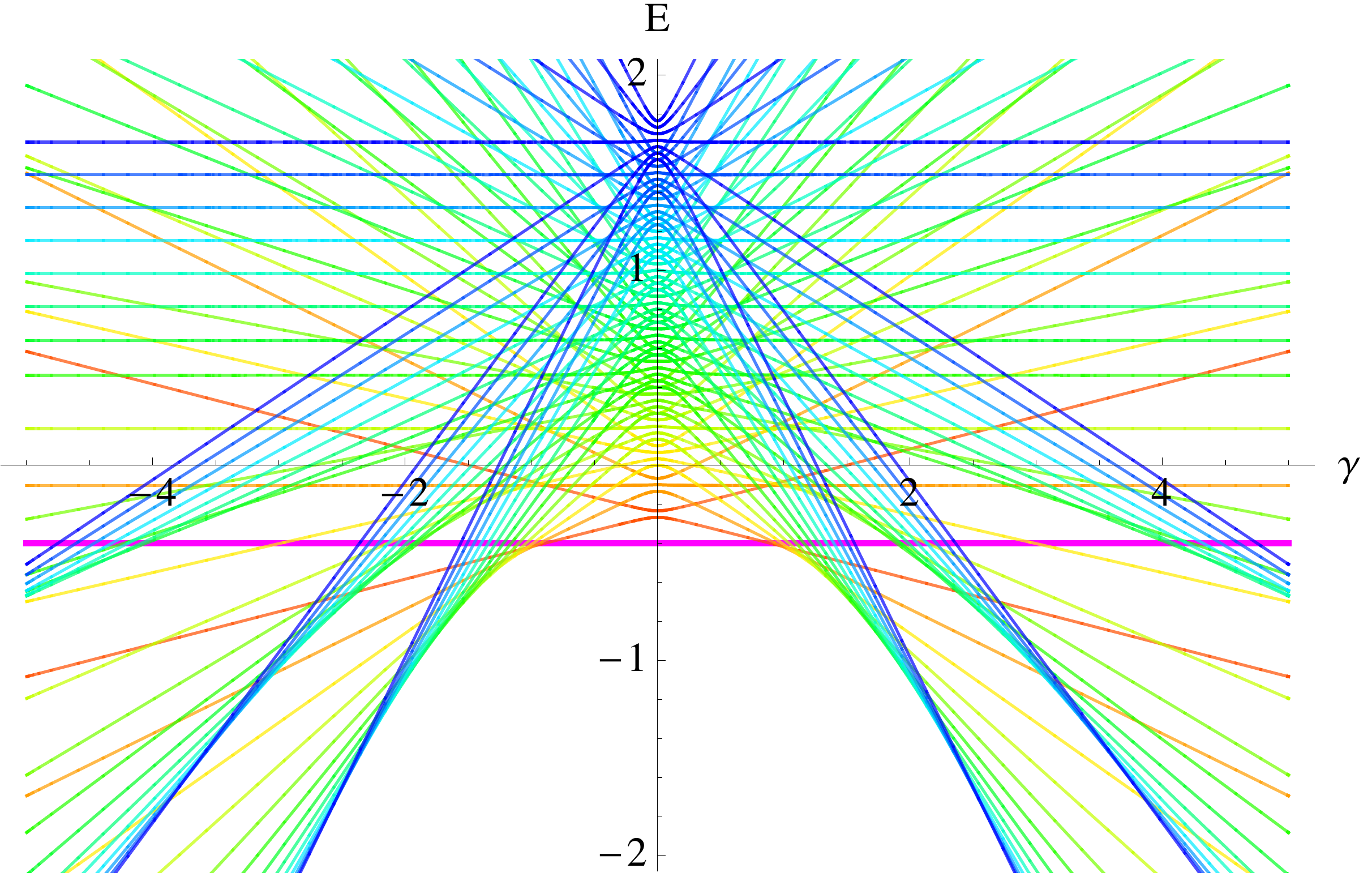}} \quad
	\scalebox{0.55}{\includegraphics{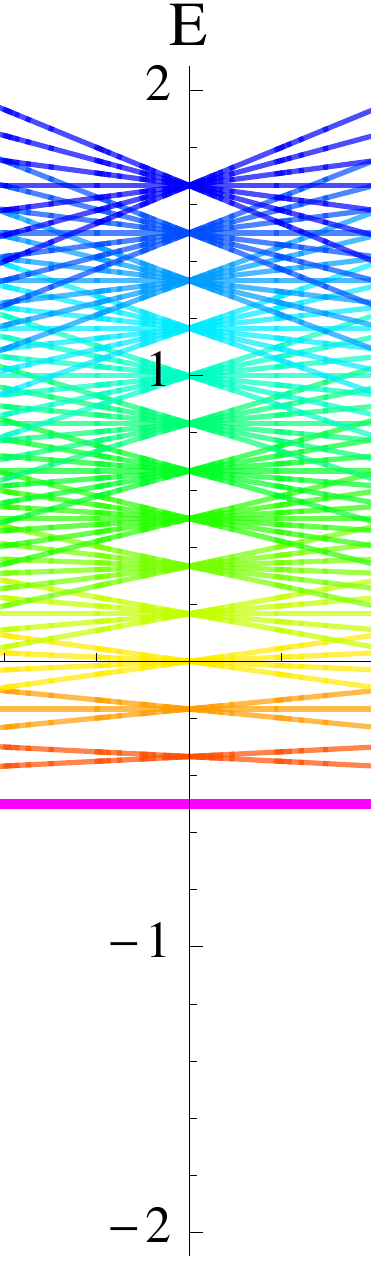}}
	\end{center}
	\caption{Colour online. Energy spectrum for the Hamiltonian (\ref{TCHamiltonian}) of the Tavis-Cummings model. $\Delta \neq 0$ (left) leads to avoided crossings. Pairs of curves of the same colour emanating from almost the same point on the energy axis correspond to the same value of $\lambda$. When $\Delta=0$ (right) curves corresponding to the same $\lambda$ touch at $\gamma=0$. The straight horizontal line at $E=-0.4$ (magenta) is the energy of the ground state in the normal region.}
\label{spectrumTCM}
\end{figure}

The eigenvectors and eigenvalues of $H$ can be obtained through diagonalisation of its associated matrix, thus allowing us to calculate the expectation value of all important field and matter observables, as well as the entanglement entropy, the squeezing parameter, and the population distributions~\cite{scrip, scrip2}. For instance, taking the natural Hilbert space basis $\vert \nu,\, j,\, m\rangle$, where $\nu$ is the eigenvalue associated to the photon number operator, $j(j+1)$ is the eigenvalue associated to the total angular momentum operator, $m$ is the particle occupation number $\vert m \vert \leq j \leq N_A/2$, where the $j=N_A/2$ holds for identical atoms. Substituting the label $m$ for the eigenvalue of the constant of motion, $\lambda = \nu + m + j$, we can obtain the full energy spectrum of $H$. This is shown in Figure~\ref{spectrumTCM} (left) for $N_A=6$ atoms, $\lambda$ up to $10$, and a detuning parameter of $\Delta=0.2$. One can see the avoided crossings due to $\Delta \neq 0$; had we $\Delta = 0$ they would touch at $\gamma = 0$ (cf. Figure~\ref{spectrumTCM} (right)). Pairs of curves of the same colour emanating from almost the same point on the energy axis (or the same one in the case for $\Delta=0$) correspond to the same value of $\lambda$. The thicker horizontal line at $E=-0.4$ (magenta) is the energy of the ground state in the normal region.

We are interested, however, in studying the system analytically. To this end, we propose to use as a {\it test-state} a direct product of coherent Heisenberg-Weyl $HW(1)$ states and $SU(2)$ states $\vert\alpha,\zeta\rangle = \vert\alpha\rangle \otimes \vert\zeta\rangle$ as:
\begin{equation}
	\fl\qquad
\vert\alpha,\zeta\rangle = \frac{\exp(-\vert\alpha\vert^2/2)}{(1+\vert\zeta\vert^2)^j}\, \sum_{\nu=0}^{\infty}\, \sum_{m=-j}^j \frac{\alpha^\nu}{\sqrt{\nu!}} \left( \begin{array}{c} 2j\\ j+m \end{array} \right)^{1/2} \zeta^{j+m}\,\, \vert\nu\rangle \otimes \vert j,\,m\rangle
\label{HWSU2states}
\end{equation}
with $\zeta = \tan\left(\frac{\theta}{2}\right) \exp(i\phi),\ \alpha = \frac{1}{\sqrt{2}}(q + ip)$, $(\theta,\phi)$ being the parameters on the Bloch sphere and $(q,p)$ the field quadratures.

The {\it energy surface}, defined as the expectation value of the Hamiltonian on the test state: $\mathcal{H} = \langle\alpha,\,\zeta\vert H \vert\alpha,\,\zeta\rangle$, is then given by
\begin{equation}
	\fl\qquad
\label{energysurface}
\mathcal{H}(q,p,\theta,\phi) = \frac{1}{2N_A}(q^2 + p^2) - \frac{1}{2}\,\omega_A\,\cos\theta + \frac{\gamma}{\sqrt{2\,N_A}}\sin\theta\, (q\cos\phi - p\sin\phi)\,.
\end{equation}

The critical points of $\mathcal{H}$ determine $3$ regions, as given by $\theta_c =0$ ({\it North Pole}), $\theta_c =\pi$ ({\it South Pole}), and $\theta_c =\arccos(\omega_A / \gamma^2)$ ({\it Parallels}); for each of these regions the minima of the energy $E_0$ and values $\lambda_c := \langle\Lambda\rangle_c$ (the expectation values of the constant of motion) are as follows:
	\begin{equation}
		\fl
		\begin{array}{llll}
		\theta_{c}=0\,,&E_{0}=-\frac{N_A\,\omega_{A}}{2}\,,&\lambda_{c}=0 \, ,&\hbox{ for }\omega_{A}>			\gamma^2\\
		\theta_{c}=\pi\,,&E_{0}=\phantom{-}\frac{N_A\,\omega_{A}}{2}\,,&\lambda_{c}=N_A \, ,
		&\hbox{ for }\omega_{A}<-\gamma^2\\
		\theta_{c}=\arccos\left(\frac{\omega_{A}}{\gamma^{2}}\right)\,,
		&E_{0}=-\frac{N_A(\omega_{A}^{2}+\gamma^{4})}{4\,\gamma^{2}}\,,&
		\lambda_{c}=\frac{N_A(-\omega_{A}\,\left(\omega_{A}+2\right)+\gamma^{4}+2\gamma^2)}{4\,\gamma^{2}} \, , &
		\hbox{ for }\left|\omega_{A}\right|<\gamma^{2} 
		\end{array}
		\label{critical}
	\end{equation}

At these critical points, $q_c = -\sqrt{N_A/2}\, \gamma \sin\theta_c \cos\phi_c,\ \hbox{and}\ p_c = \sqrt{N_A/2}\, \gamma \sin\theta_c \sin\phi_c$, so that matter and field variables combine. As $\phi$ is a cyclic variable, $\phi_c$ may be taken arbitrarily. We set $\phi_c=0$, and the expressions in terms of this variable may be recovered by performing a rotation through an angle $\phi$ around the $z$-axis in the appropriate phase space: $(q,\,p)$ and $(J_x,\,J_y)$ for field and matter quantities respectively.

We can write explicitly the form that the states take in each of these $3$ regions:
\begin{equation}
	\fl
	\begin{array}{llll}
	\hbox{North Pole:} & \omega_{A}>\gamma^2 & \quad \vert\psi_{np}\rangle = \vert 0 \rangle \,\otimes\, \vert j,\, -j\rangle \\
	\hbox{South Pole:} & \omega_{A}<-\gamma^2 & \quad \vert\psi_{sp}\rangle = \vert 0 \rangle \,\otimes\, \vert j,\, j\rangle \\
	\hbox{Parallels:} & \vert\omega_{A}\vert < \gamma^2 & \quad \vert\psi_{par}\rangle = \sum_{m=-j}^{+j}\, \sum_{\nu=0}^{+\infty} A_{m,\nu}\, \vert \nu \rangle \,\otimes\, \vert j,\, m\rangle \\ 
	\end{array}
	\label{states3regions}
\end{equation}
with
\begin{eqnarray*}
A_{m,\nu} = &{2j \choose j+m}^{1/2} \exp\{-\frac{j\gamma^2}{4} \left( 1-\frac{\omega_A^2}{\gamma^4} \right)\}\, \frac{(-\sqrt{2j}\,\gamma)^\nu}{\sqrt{\nu!}} \\
&\times \left( \frac{1}{2} + \frac{\omega_A}{2\gamma^2} \right)^{(j-m+\nu)/2} \left( \frac{1}{2} - \frac{\omega_A}{2\gamma^2} \right)^{(j+m+\nu)/2}\,.
\end{eqnarray*}

\begin{figure}
		\begin{center}
		\scalebox{0.5}{\includegraphics{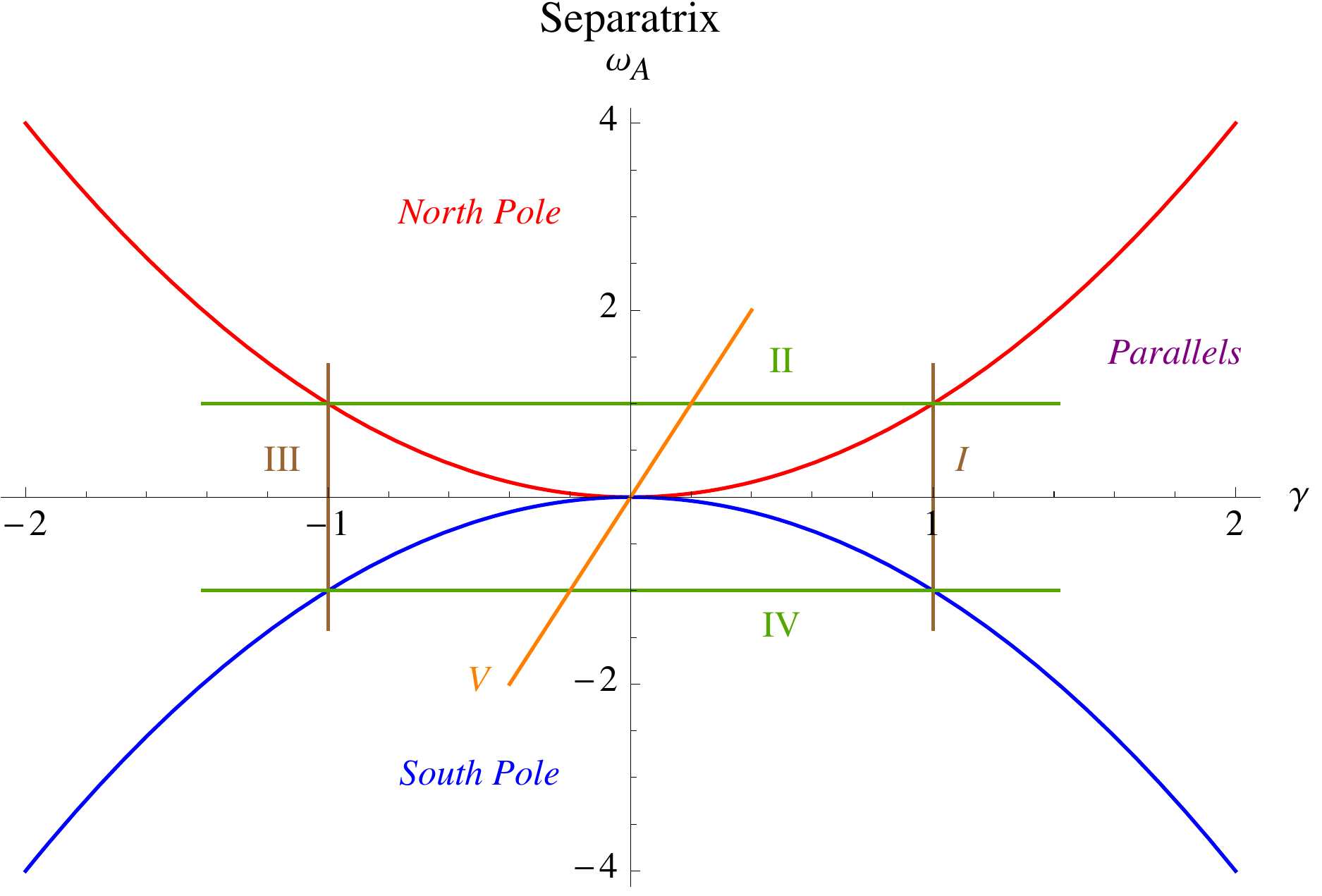}}
		\end{center}
		\caption{Colour online. Phase diagram for the Tavis-Cummings model. The normal region is described by the North and South Poles, separated by parabolae from the Parallels which denote the collective region. These parabolae constitute the {\it separatrix}. Crossings along paths $I,\ II,\ III$, and $IV$ (horizontal (green) and vertical (brown) straight lines) lead to second-order phase transitions; only the crossing $V$ (slanted (orange) straight line) through the origin gives a first-order phase transition.}
	\label{SeparatrixTCM}
	\end{figure}

The $3$ regions define also a {\it separatrix}, where the Hessian of $\mathcal{H}$ is singular.  This is given by $\omega_A = \pm \gamma^2$, and is shown in Fig.~\ref{SeparatrixTCM}. Crossing the separatrix along paths $I,\ II,\ III$, and $IV$ (horizontal (green) and vertical (brown) straight lines in the figure) leads to second-order phase transitions; crossing it along path V to first order transitions.

\begin{figure}
		\begin{center}
		\scalebox{0.35}{\includegraphics{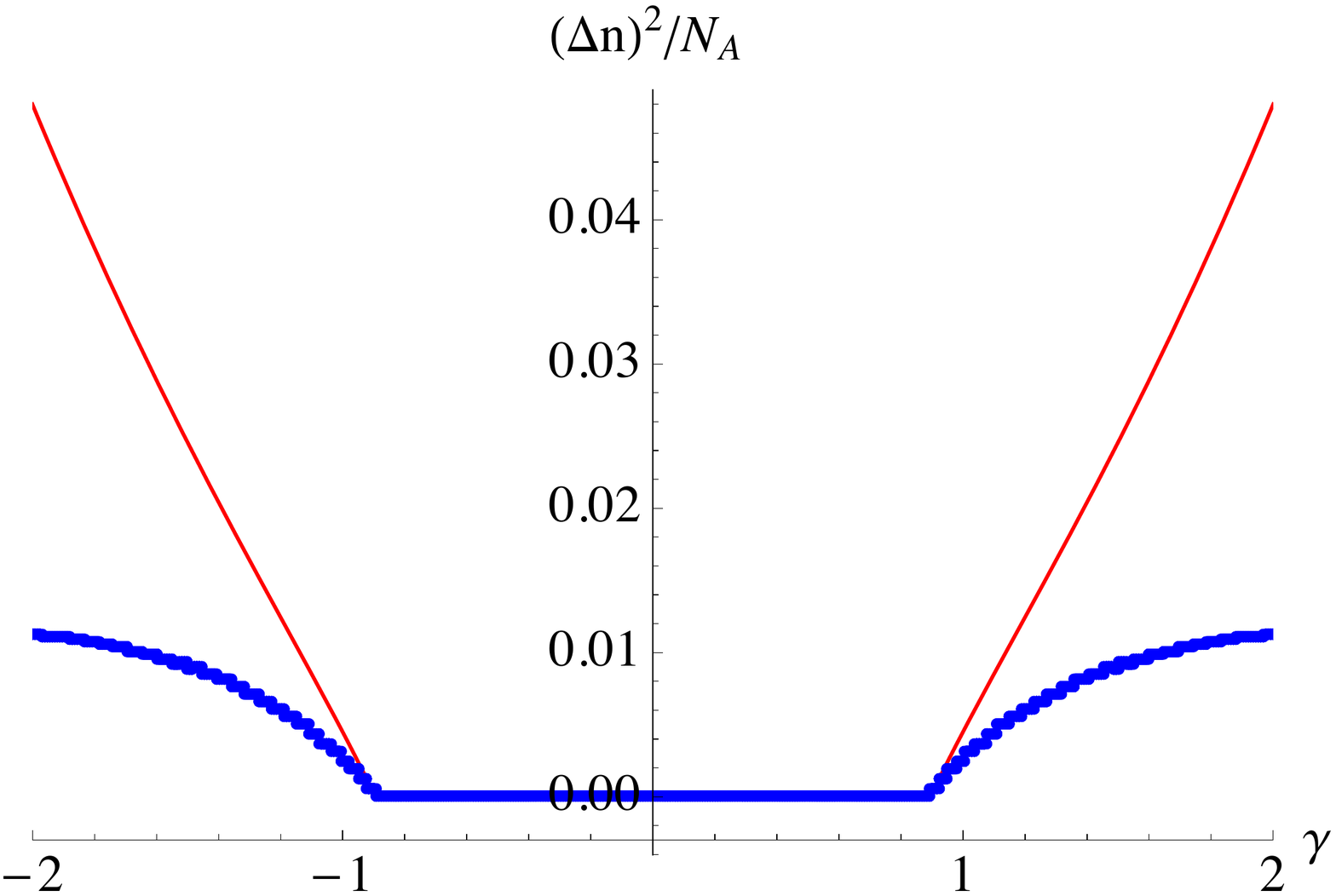}} \quad
		\scalebox{0.35}{\includegraphics{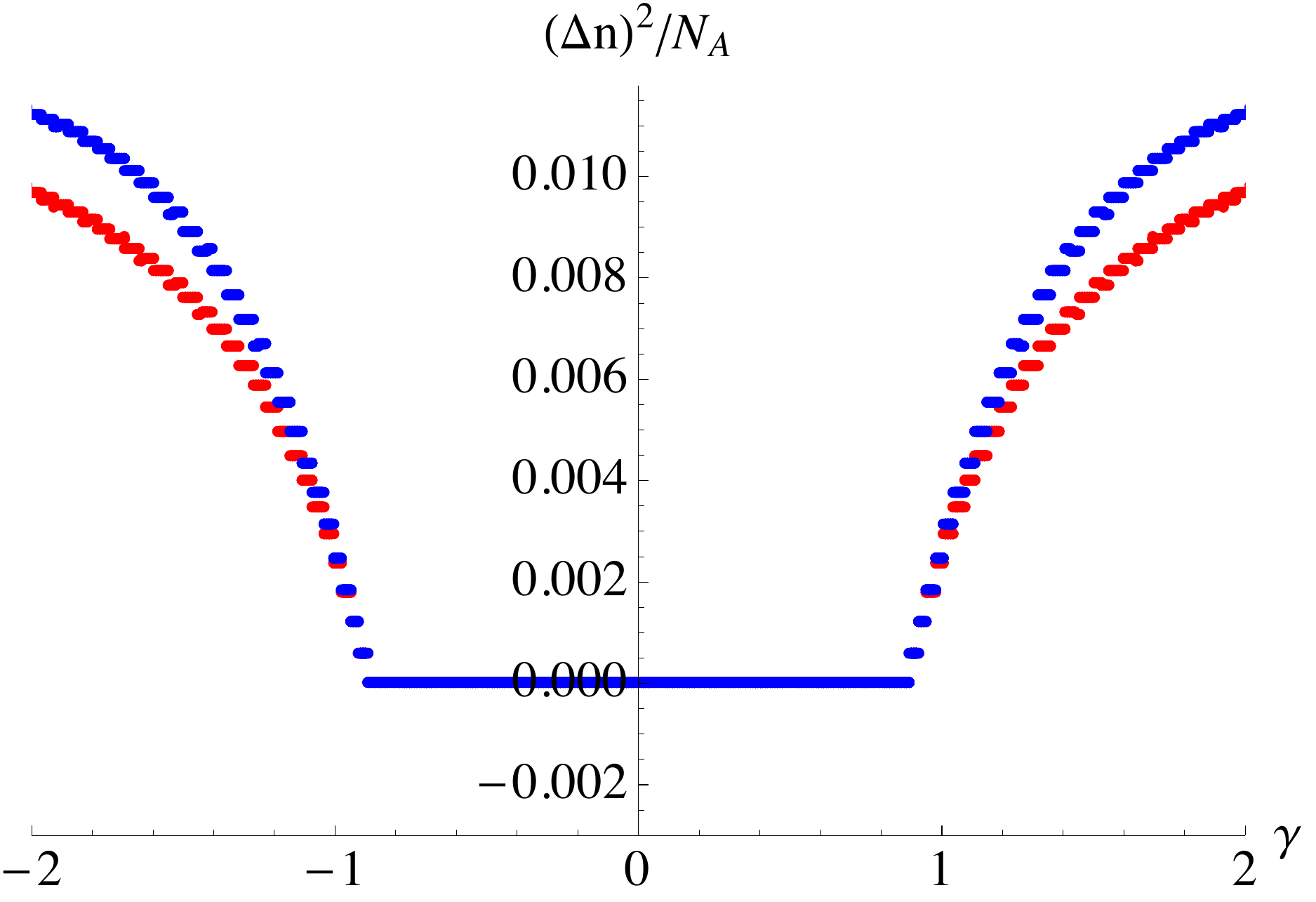}}
		\end{center}
		\caption{Colour online. Left: Comparison between the fluctuation in the expectation value of the number of photons $(\Delta n)^2/N_A$ given by the coherent state approximation (continuous, red curve) and the exact quantum solution (discontinuous, blue curve), shown as functions of $\gamma$. Right: The same comparison using projected states (continuous, red line); note the scale in the ordinate axis. We have used $N_A=20$, and $\Delta=0.2$ in both graphs. The noticeable differences arise from the fact that the coherent state contains contributions from all eigenvalues $\lambda = \nu + m + j$ of $\hat\Lambda$, and therefore does not reflect the symmetry of the Hamiltonian.}
	\label{nf_TCM}
	\end{figure}

In general, these coherent variational states approximate very well the properties of the ground state of the quantum solution~\cite{scrip}. This is true for the energy, the constant of motion $\lambda(\gamma)$, and the matter observables $\langle J_z\rangle$ and its fluctuation squared $(\Delta J_z)^2$, etc. Even the expectation value of the number of photons $n = \langle \hat{N}_{ph} \rangle$ is well approximated; but its fluctuation, as well as other properties of the system such as the occupation probabilities, are not: Fig.~\ref{nf_TCM} (left) shows how bad an approximation to the photon number fluctuation we get. The noticeable differences arise from the fact that the coherent state contains contributions from all eigenvalues $\lambda = \nu + m + j$ of $\Lambda$, and therefore does not reflect the symmetry of the Hamiltonian leading to the constant of motion.

One may maintain the symmetry through a projection of the variational tensorial product of coherent states onto the value of the constant of motion of the TCM which minimises the (classical) energy of the ground state.  This projection restores the Hamiltonian symmetry and is amiable to analytical calculations. 

Projecting, the state becomes
\begin{equation}
	\fl
	\label{projectedstate}
	\qquad \vert \psi \rangle = \mathcal{N}
		\left\{
			\begin{array}{ll}
			\vert 0\rangle\otimes\vert j,\,-j\rangle \ ,
			&\ \omega_{A}>\gamma^{2}\\
			\sum_{\nu=\max[0,\,\lambda-2j]}^{\lambda}\,
			{2j \choose \lambda - \nu}^{1/2}
			\,\frac{\eta^{\nu}}{\sqrt{\nu!}}\,\,\vert\nu
			\rangle\otimes\vert j,\,\lambda-j-\nu\rangle\ ,
			&\left|\omega_{A}\right|\leq \gamma^{2}\\
			\vert 0\rangle\otimes\vert j,\,j\rangle\ ,
			&\ \omega_{A}<-\gamma^{2}
			\end{array}
		\right.
\end{equation}
where we have defined $\eta = - \frac{\sqrt{N_A}\gamma}{2} (1+\frac{\omega_A}{\gamma^2})$. The factor $\mathcal{N}$ is the normalisation factor. With respect to these projected states, the energy surface is given in terms of associated Laguerre polynomials~\cite{scrip2} as follows
\begin{eqnarray}
\label{energysurface_proj}
\fl
&\mathcal{H} =& \frac{\lambda - j + j\Delta}{2j} - \left[ \Delta - \frac{2\gamma}{\sqrt{2j}}\eta\right] 
	\left\{
		\begin{array}{ll}
		L^{2j-\lambda}_{\lambda-1}(-\eta^2) / L^{2j-\lambda}_{\lambda}(-\eta^2) \ ,
		& 1 \leq \lambda \leq 2j \\
		\frac{\lambda}{2j} L^{\lambda-2j}_{2j-1}(-\eta^2) / L^{\lambda-2j}_{2j}(-\eta^2)\ ,
		&\ \ \ \ \ \ \lambda \geq 2j
		\end{array}
	\right.
	\\
\fl&\mathcal{H} =& -\frac{1}{2}(1-\Delta)\ , \hspace*{3.6\mathindent}\lambda=0
\end{eqnarray}
and the approximation to the photon number fluctuation is restored as shown in Fig.~\ref{nf_TCM} (right) (note the scale of the ordinate axis).

\begin{figure}[h]
	\begin{center}
\scalebox{1}{\includegraphics{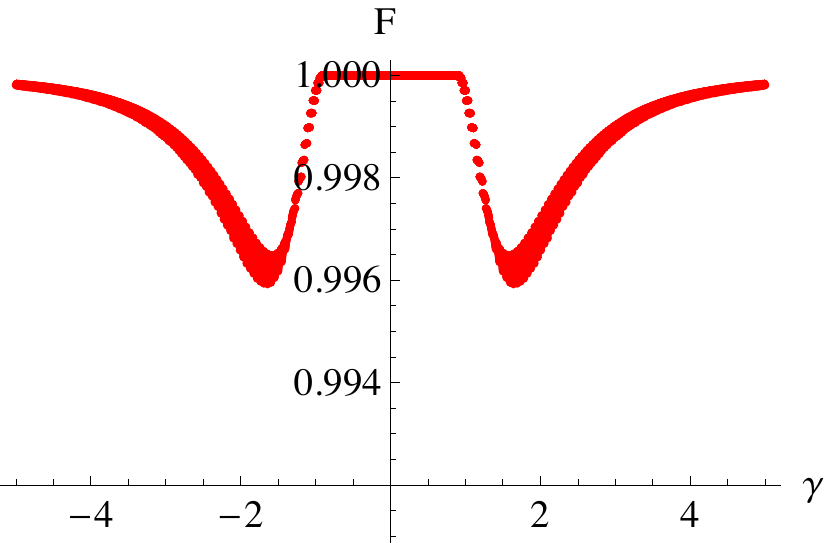}}
\caption{Fidelity $F$ between the projected state and exact quantum ground states, as a function of the interaction strength $\gamma$. The plot is for a detuning parameter $\Delta=0.2$ and $N_A = 20$ atoms.}
	\end{center}
\label{fidelity_TCM}
\end{figure}

A better way to measure the ``distance'' between states is via the fidelity,

	\[
		F\left(\varrho_{1},\,\varrho_{2}\right)=
		\,\hbox{tr}\left(\sqrt{\sqrt{\varrho_{1}}\,
		\varrho_{2}\,\sqrt{\varrho_{1}}}\right)\ ,
	\]
where $\varrho_{1}$ and $\varrho_{2}$ denote the density matrices of the states in question. For pure states, this definition coincides with the square of the scalar product between the states~\cite{zanardi06}. Figure~\ref{fidelity_TCM} shows a perfect overlap $F = 1$ between the projected and quantum states in the normal region, dropping to $F = 0.996$ when crossing the separatrix into the Parallels region, only to recover again towards $F = 1$ as $\gamma$ grows.

Even if our approximation by projected (symmetry-adapted) states is not exact, an excellent approximation to the exact quantum solution of the ground state of the TCM model is obtained. What is gained is that these states have an analytical form in terms of the model parameters and allow for the analytical calculation of the expectation values of field and matter observables, as well as for the study of the phase diagram of the system.

\subsection{The Dicke Model}

When the RWA approximation is {\it not} taken, we have the full Dicke Hamiltonian given in \eref{DickeHamiltonian}. Once again, one may obtain analytical expressions for the energy and expectation values of the relevant operators of the system via the use of the Heisenberg-Weyl and $SU(2)$ coherent states (\ref{HWSU2states}) as trial states, and the variational procedure described above. This trial state contains $N=2j$ particles distributed in all the possible ways between the two levels and up to an infinite number of photons in the cavity. The energy surface in this case is given by
\begin{equation}
	\fl \qquad
	\mathcal{H}(q,p,\theta,\phi) = \frac{1}{2N_A}(q^2 + p^2) - \frac{1}{2}\,\omega_A\,\cos\theta + \frac{\sqrt{2}\,\gamma}{\sqrt{N_A}} q\,\sin\theta\,\cos\phi \, ,
	\label{energysurfaceDicke}
\end{equation}
and the separatrix shrinks to $\omega_A = 4\gamma_c^2$, for $\gamma_c$ the critical value of $\gamma$. As before, the crossings of this separatrix are second-order phase transitions, except for the first-order crossing through the origin. The energy minima in the normal and collective (superradiant) regions are
\begin{eqnarray}
	E_{normal} &=& -2N_A\gamma_c^2 \nonumber\\
	E_{superradiant} &=& -N_A\gamma^2 \left[\left(\frac{\gamma_c}{\gamma}\right)^4 + 1 \right] \, ,
	\label{energyminimaDicke}
\end{eqnarray}
and the expected number of photons are
\begin{eqnarray}
	\langle\hat{N}_{ph}\rangle_{normal} &=& 0 \nonumber\\
	\langle\hat{N}_{ph}\rangle_{superradiant} &=& N_A\gamma^2 \left[1 - \left(\frac{\gamma_c}{\gamma}\right)^4\right] \, ,
	\label{photonsDicke}
\end{eqnarray}
which calls for the definition of $x = \gamma/\gamma_c$.

\begin{figure}
		\begin{center}
		\scalebox{0.35}{\includegraphics{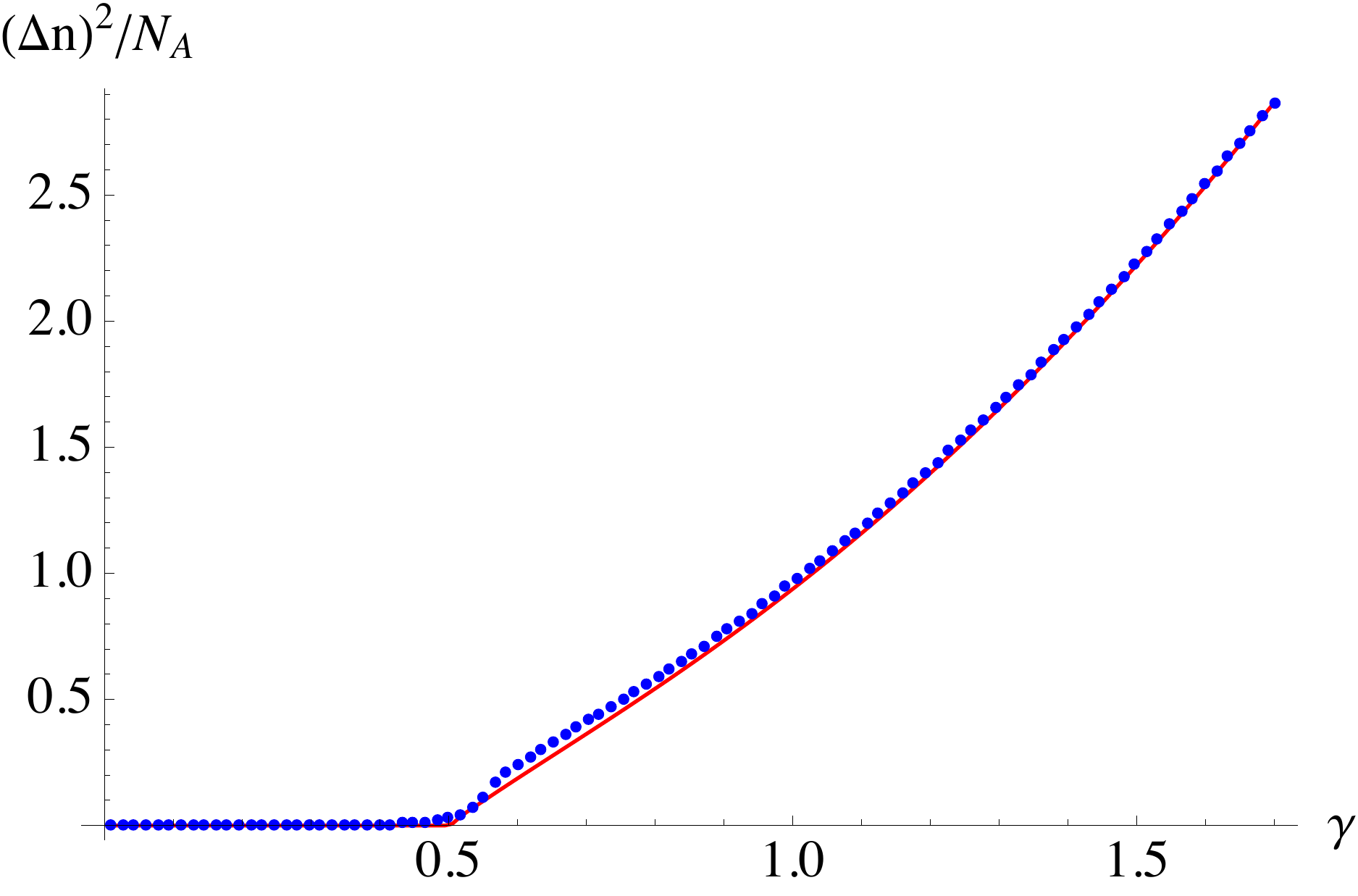}} \qquad
		\scalebox{0.35}{\includegraphics{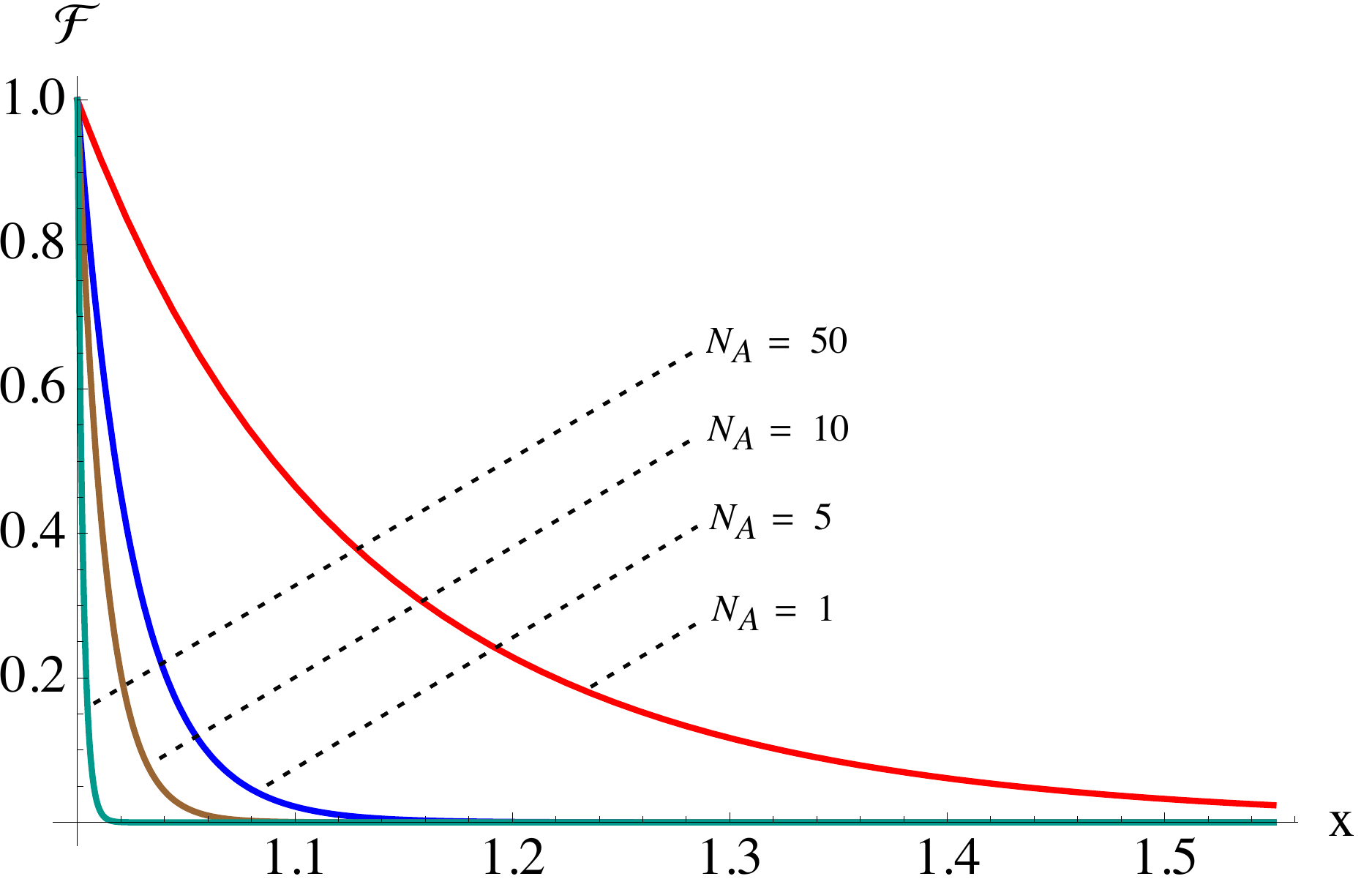}}
		\end{center}
		\caption{Colour online. Left: comparison between the fluctuation in the expectation value of the number of photons $(\Delta n)^2/N_A$ given by the symmetry-adapted state approximation (continuous, red curve) and the exact quantum solution (discontinuous, blue curve), shown as functions of $\gamma$. We have used $N_A=20$, and $\Delta=0$. Right: $\mathcal{F}$ functional dependance on $x$ for different values of $N_A$. As $N_A$ increases $\mathcal{F}$ tends to zero much more rapidly.}
	\label{nf_Dicke}
	\end{figure}

As $\hat\Lambda = \sqrt{\hat{J}^2 + 1/4} -1/2 + \hat{J}_z + a^{\dagger}a$ is no longer a constant of motion for the system we cannot simply project onto one of its eigenvalues, rather, we have a dynamical symmetry associated with the projectors of the symmetric and antisymmetric representations of the cyclic group $C_2$, given by
\begin{equation}
	P_{\pm}= \frac{1}{2} \left( 1 \pm e^{i \pi \hat{\Lambda}} \right) \, .
	\label{projectorsDicke}
\end{equation}
This symmetry allows, however, for the classification of the eigenstates in terms of the parity of the eigenvalues $\lambda = j +m + \nu$ of $\hat\Lambda$~\cite{papercorto}. Adapting the coherent states to the parity symmetry of the Hamiltonian then amounts to sum over $\lambda$ even or odd, with two resulting orthogonal states $\vert\alpha,\,\zeta,\,\pm\rangle$. For these states the energy surface associated to the superradiant regime takes the form
\begin{equation}
	\langle H \rangle_{\pm}=
      -N\gamma_c^2 x^2\,\left[ 2 - (1-x^{-4})\,\frac{1\mp \mathcal{F}}{1\pm \mathcal{F}}\right]\ ,
      \label{symad2}
\end{equation}
with
\begin{equation}
      \mathcal{F}=x^{-2N_A}\,\e^{-2N_A\,\gamma_c^{2}\,x^2\left(1-x^{-4}\right)}\ ,
\end{equation}
and the limit $x\rightarrow 1$ gives the expressions for the normal region. The fidelity between these {\it symmetry-adapted} states and the exact quantum states is very close to $1$ except in a small vicinity of the transition region in phase space, so it is no surprise that they provide an excellent agreement with the expectation values of the quantum operators for the system, an example of which is shown in Fig.~\ref{nf_Dicke} (left) for the fluctuation in the expectation value of the number of photons $(\Delta n)^2/N_A$ as given by our projected state approximation (continuous, red curve) compared with the exact quantum solution (discontinuous, blue curve), as functions of $\gamma$. We have used $N_A=20$, and the resonant condition $\Delta=0$.

If we calculate the overlap between the coherent and adapted states we obtain
\begin{equation}
	\vert \langle\alpha_c\,\zeta_c\,\vert\,\alpha_c\,\zeta_c,\,\pm\rangle\,\vert^2 = \frac{1}{2}\,(1 \pm \mathcal{F})
	\label{overlapcohsym}
\end{equation}
and, since the behaviour of $\mathcal{F}$ falls very rapidly with $\gamma$ (cf. Fig.~\ref{nf_Dicke} (right)), this overlap will be at best equal to $1/2$, which makes the ordinary coherent states a good approximation only in special cases. \ref{appA} compares the expectation values and fluctuations of matter and field observables for the coherent and symmetry-adapted states, evaluated at the critical points for the energy surface (\ref{energysurfaceDicke}). For expectation values different from zero in the symmetry-adapted states, the coherent state results can be obtained from the former by letting $\mathcal{F}$ go to zero. Notable exceptions are the field quadratures $(q,\,p)$ and the atomic operator fluctuations. For large $N_A$ the function $\mathcal{F}$ tends to zero even more rapidly; this is why coherent states have been so successful in the past as trial functions.

Like the quantum states, the symmetry-adapted states show no divergences for field or matter expectation values at the phase transition. This is in contrast with results found previously~\cite{nagy10, emary03, lambert04}, which are an artifact of an inappropriate truncation of the Hamiltonian. For more good properties of the symmetry-adapted states, including probability distributions of photons, of excited atoms, and their joint distribution, cf.~\cite{paperextenso}. In particular, even though the coherent states, the symmetry-adapted states, and the quantum states, are quantities arrived at via very different methods, they show a universal character in that a universal parametric curve for {\it any} number of atoms $N_A$ is obtained for the first quadrature of the electromagnetic field, $q$, and for the atomic relative population $\langle J_z \rangle$, as implicit functions of the atom-field coupling parameter $\gamma$, valid for both the ground- and first-excited states~\cite{universal}. Furthermore, for {\it all} values of the coupling parameter and again {\it any} number of atoms, the behaviour of the number of photons vs. the relative atomic population is universal.

\subsubsection{Critical Exponents}

For a homogeneous function $f(r)$ we have $f(\beta r) = g(\beta)\,f(r)$ for all values of $\beta$. The scaling function $g(\beta)$ is of the form $g(\beta) = \beta^s$; $s$ is called the {\it critical exponent}. It is known that the singular part of many potentials in physics are homogeneous functions near second-order phase transitions; in particular, this is true for all thermodynamic potentials~\cite{katzgraber}. The behaviour of important observables of a system near phase transitions may thus be described by the system's critical exponent, and these are believed to be universal with respect to physical systems.

\begin{figure}
		\begin{center}
		\scalebox{0.35}{\includegraphics{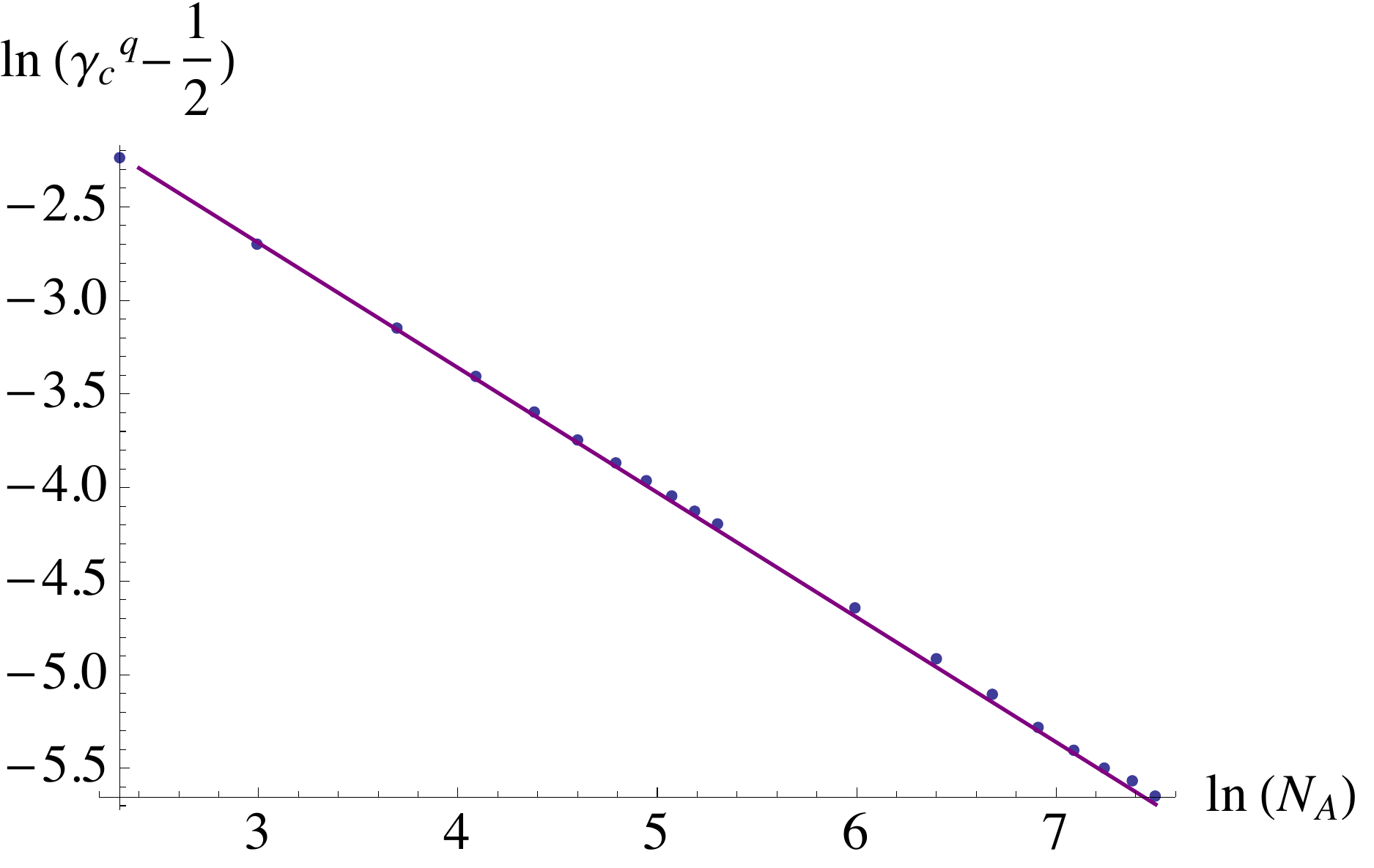}} \qquad
		\scalebox{0.35}{\includegraphics{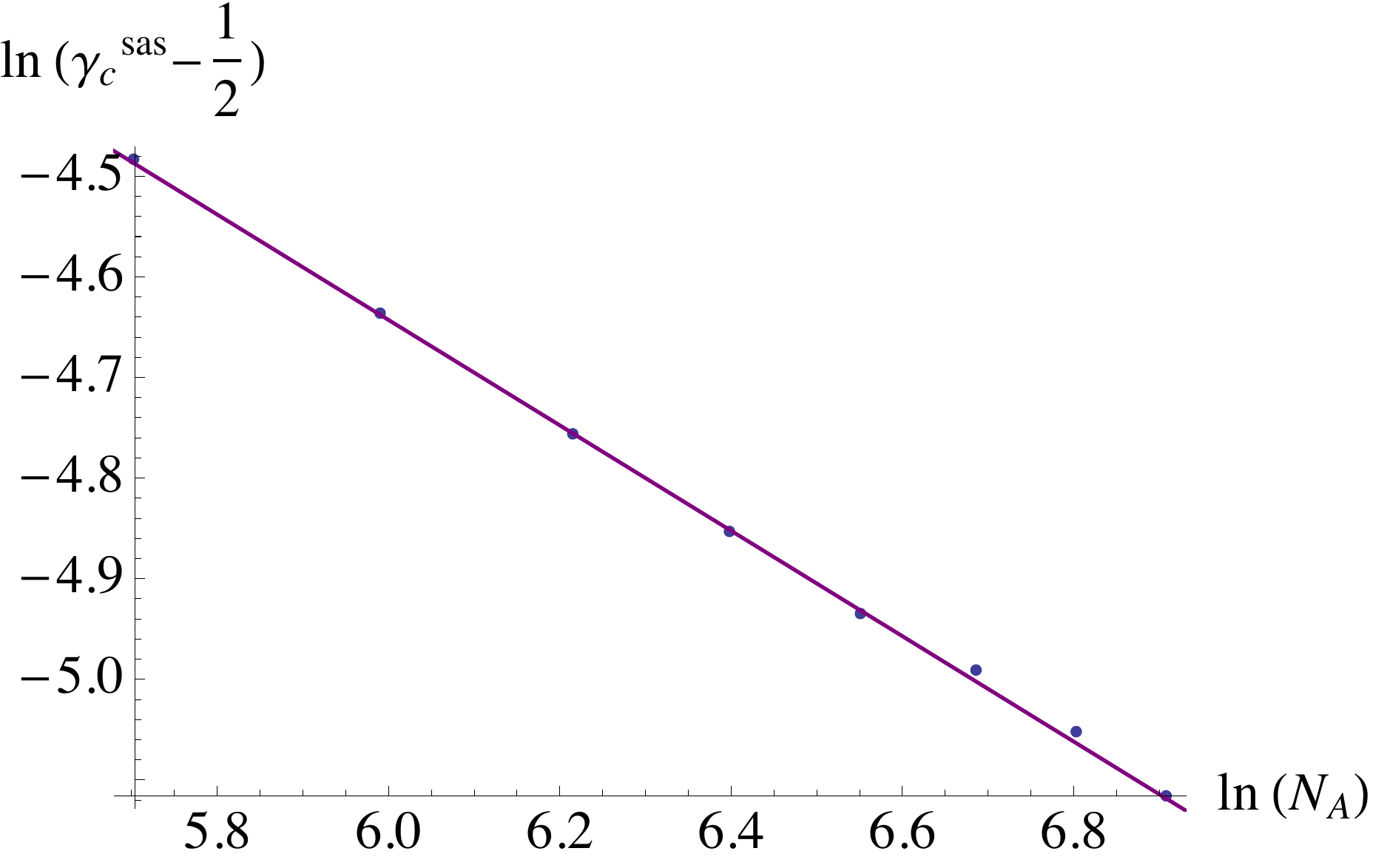}}
		\end{center}
		\caption{Logarithmic behaviour of the critical value of the coupling parameter $\gamma_c$ with the number of atoms $N_A$. For the quantum (q) states the critical exponent is $-2/3$, while for the symmetry-adapted states (SAS) it turns out to be $-11/21$.}
	\label{crit_exp_Dicke}
	\end{figure}

Our treatment for {\it finite} $2$-level systems in a cavity, in the presence of a radiation field, allows us to study the critical value of the atom-field coupling parameter $\gamma_{c}$ as a function of the number of atoms $N_A$, from which its critical exponent may be derived. Figure~\ref{crit_exp_Dicke} shows this relationship for the ground state of both the quantum states (left) and the symmetry-adapted states (SAS) (right). For the quantum states the points correspond to a numerical solution from diagonalising the Hamiltonian, and the continuous curve to a model fit. The value of $\gamma_{c}$ was obtained by calculating in parameter space the place where the fidelity function between neighbouring states (cf. \eref{fidelity} below) vanishes. We varied $N_A$ from $10$ to $1800$ and the logarithm of the variables is plotted for a more demanding fit, obtaining
\begin{equation}
	\ln \left(\gamma_c^q - \frac{1}{2}\right) = \ln \left(\frac{1}{2}\right) - \frac{2}{3}\,\ln\left(N_A\right)\, ,
\end{equation}
or, equivalently,
\begin{equation}
	\gamma_c^q = \frac{1}{2} + \frac{1}{2}\, N_A^{-\frac{2}{3}}\, .
\end{equation}

Except for a small vicinity of the phase transition, the SAS states do approximate very well the quantum solutions. However, the critical exponent obtained for the asymptotic behaviour of the adapted states is $-11/21$, as opposed to $-2/3$, as shown in the figure (right). This is precisely because the evaluation takes place at the phase transition point, where the states (quantum and adapted) differ most~\cite{camop2}. The value of $\gamma_{c}$ for the SAS states was obtained by calculating in parameter space the place where the minimum of the energy $E_{+\,min}$ for the state $\vert \alpha_c\,\zeta_c,\,+\rangle$ presents a discontinuity. Since we are interested in the asymptotic behaviour, we took $N_A$ from $200$ to $1000$; the continuous curve shows the fit
\begin{equation}
	\gamma_c^{sas} = \frac{1}{2} + \e^{-\frac{1}{2}}\, N_A^{-\frac{11}{21}}\, .
	\label{ExpCritSAS}
\end{equation}
%

Table~\ref{tgammavsN} shows a sample of values of $(N_A,\,\gamma_c)$ for the quantum and the SAS ground states, in order to make explicit the fact that for small $N_A$ the values of the quantum critical interaction parameter $\gamma_c^q$ differ considerably from that of the SAS states $\gamma_c^{sas}$. This difference tends to zero as $N_A$ increases, and in the limit $N_A\to\infty$ the phase transition region in phase space coincides for both states at $\gamma_c=0.5$.

\begin{table}
\caption{Sample of values of the quantum critical interaction parameter $\gamma_c^q$ and the SAS critical interaction parameter $\gamma_c^{sas}$, for different values of $N_A$.}
\begin{center}
\begin{tabular}{c|cc}
$N_A$ & $\gamma_c^q$ & $\gamma_c^{sas}$\\
\hline  & & \\[-3mm]
$20$ & 0.5677 & 0.5522 \\
$40$ & 0.5432 & 0.5343 \\
$100$ & 0.5236 & 0.5204 \\
$400$ & 0.5096 & 0.5097 \\
$800$ & 0.5061 & 0.5068 \\
$1000$ & 0.5051 & 0.5060 \\
\end{tabular} 
\end{center}
\label{tgammavsN}
\end{table}

\begin{figure}
		\begin{center}
		\scalebox{0.25}{\includegraphics{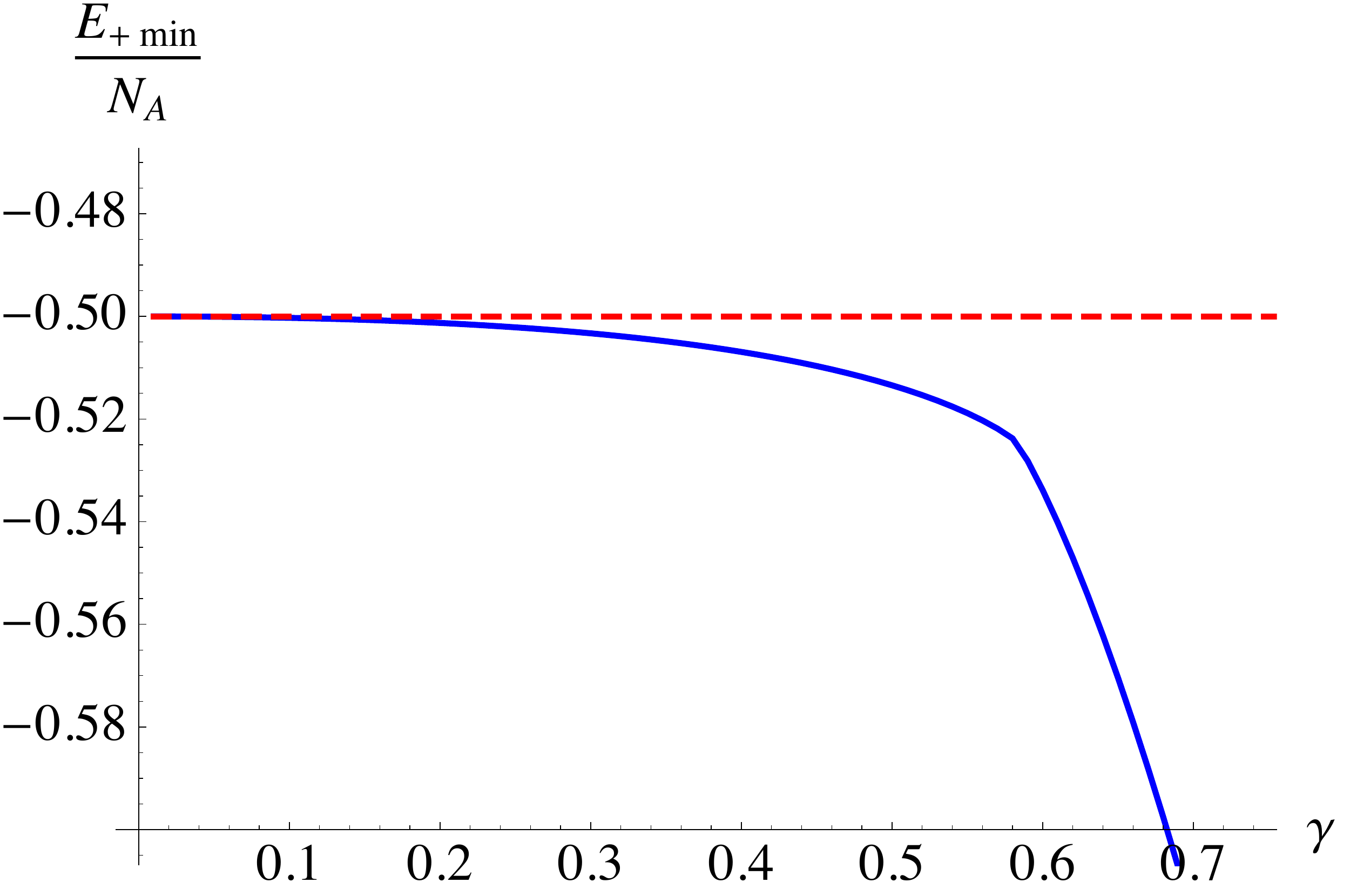}} \qquad
		\scalebox{0.25}{\includegraphics{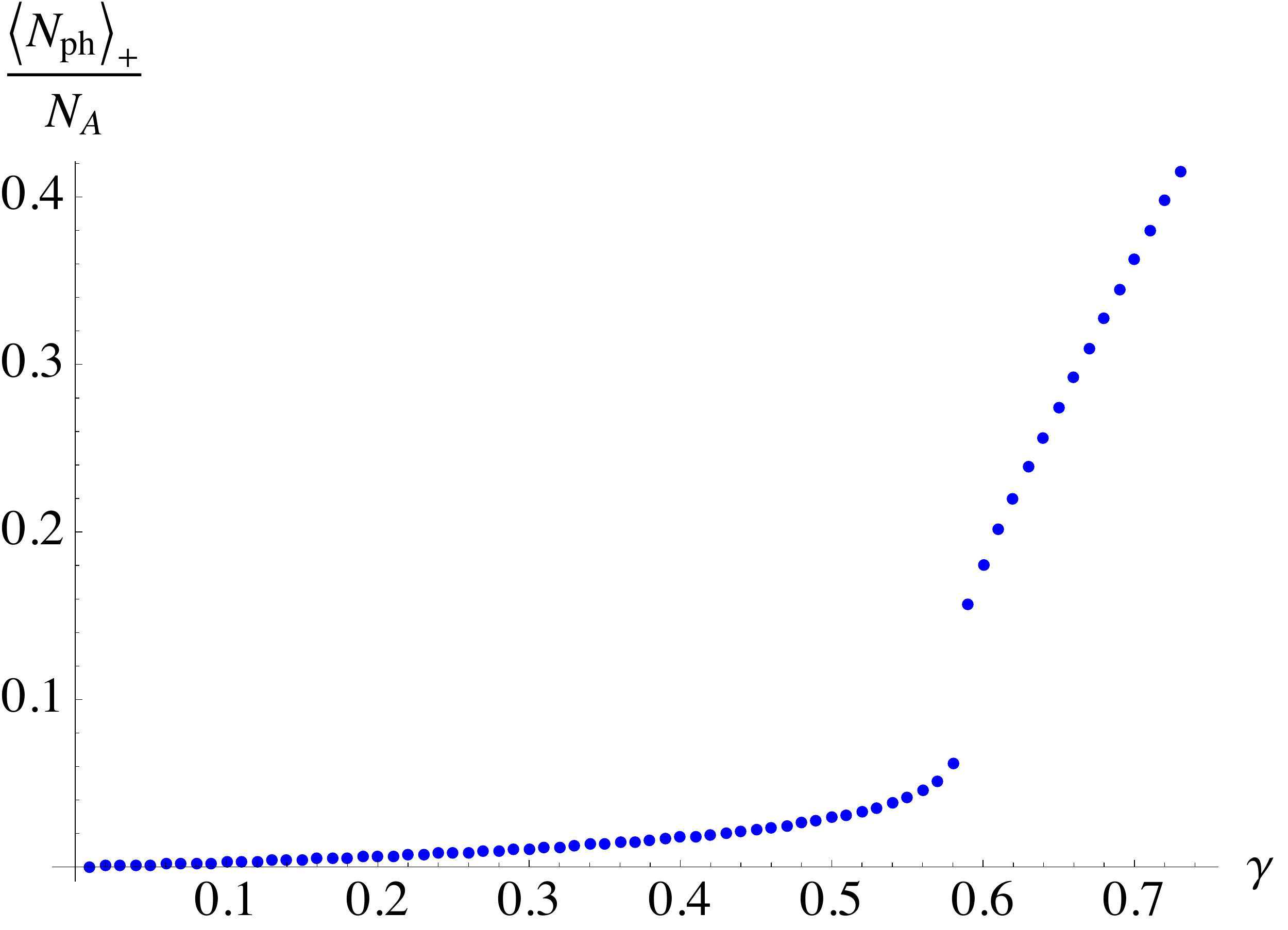}}
		\end{center}
		\caption{Energy per particle (left) and the expected number of photons per particle (right) for the symmetry-adapted ground state inside the normal region. The (red, dashed) straight line at $E_{+\,min}/N_A = -0.5$ is a comparative reference for the energy of the ground coherent state. We have taken $N_A=10$.}
	\label{NormalSAS}
	\end{figure}

It is interesting to note, from \eref{overlapcohsym}, that only for $\gamma=0$ (i.e., no matter-field interaction) do we have 
\begin{equation}
	\vert \langle\alpha_c\,\zeta_c\,\vert\,\alpha_c\,\zeta_c,\,\pm\rangle\,\vert^2 = 1 \qquad (\gamma=0)\, ,
\end{equation}
i.e., the overlap between coherent and symmetry-adapted states is perfect only when the interaction Hamiltonian $H_{int}$ vanishes. As soon as there is an interaction, no matter how small, the states differ. This is due to the fact that the ground states coincide only at $\gamma=0$. Even in the normal regime, where the coherent ground state has exactly zero photons, the SAS ground state (just as the quantum ground state) is a superposition of states with an expectation value for $N_{ph}$ different from zero. This is true for any finite number of atoms $N_A$. Figure~\ref{NormalSAS} shows the energy per particle and the expected number of photons per particle for the ground symmetry-adapted state inside the normal region. We have taken $N_A=10$ to make the distinction visually clear.

In the asymptotic limit $x\to\infty$, \eref{overlapcohsym} gives a value of $1/2$ for the overlap of the coherent and the SAS ground states. The same is true in the limit $N_A\to\infty$. This is to be expected, as the SAS ground state has contributions only from the even-parity components of the coherent ground state.

\section{Three-level Systems}

	A $3$-level system of $N_A$ atoms interacting dipolarly with an electromagnetic field of frequency $\Omega$ is described by the intrinsic Hamiltonian given in equations (\ref{Hamiltonian3Levels}, \ref{HD}, \ref{Hint}). Once again, we may take $\Omega=1$ and measure all frequencies in units of the field frequency. As mentioned before, the $i$-th level atomic frequency is denoted by $\omega_i$ with the convention $\omega_1\leq\omega_2\leq\omega_3$, and the coupling parameter between levels $i$ and $j$ is $\mu_{ij}$. The three different atomic configurations are chosen by taking the appropriate value $\mu_{ij}=0$ (cf. Fig.~\ref{configurations}). It is also convenient to define a {\it detuning parameter} $\Delta_{ij} = \omega_i - \omega_j -\Omega$ between levels $i$ and $j$.
	
In the RWA approximation the Hamiltonian reduces to~\cite{LAOP}
\begin{eqnarray}
	\label{Ham3LevelsRWA}
	\fl
H &=& \Omega\, a^\dag a + \omega_1\,A_{11} + \omega_2\,A_{22} + \omega_3\,A_{33}  \\
\fl
&&\ -\frac{1}{\sqrt{N_A}} \left[\mu_{12}\left(a\,A_{21} + a^\dag\,A_{12} \right) + \mu_{13}\left(a\,A_{31} + a^\dag\,A_{13} \right)
+ \mu_{23}\left(a\,A_{32} + a^\dag\,A_{23}\right) \right] \nonumber
\end{eqnarray}
and it has $2$ constants of motion, viz., the total number of atoms $N_A = \sum_{i=1}^3 A_{ii}$, and the total number of excitations $M = a^\dag a + \lambda_2 A_{22} + \lambda_3 A_{33}$, where the value of $\lambda_i\ (i=2,3)$ depends on the configuration taken (cf. Table~\ref{tconfig}).

\begin{table}[h!]
\caption{Values of $\lambda_{i}$, $i=2,\,3$, for the constant of motion $M$ in the different configurations.}
\begin{center}
\begin{tabular}{c|cc}
Configuration & $\lambda_{2}$ & $\lambda_{3}$\\
\hline  & &\\[-3mm]
$\Xi$ & 1 & 2\\
$\Lambda$ & 0 & 1\\
$V$ & 1 & 1\\
\end{tabular} 
\end{center}
\label{tconfig}
\end{table}

Notice that the Hamiltonian~(\ref{Ham3LevelsRWA}) is invariant under the transformation $a\to -a$ and $a^\dagger\to -a^\dagger$, which preserves the commutation relations of the bosonic operators. For this reason we consider only positive values for $\mu_{ij}$. As the system cannot be solved analytically, one may solve via numerical diagonalization. A natural basis in which we diagonalize our Hamiltonian is $\vert\nu;\,q,\,r\rangle$~\cite{cordero1}. Here, $\nu$ represents the number of photons of the Fock state; $r,\, q-r$ and $N_A-q$ are the atomic population of levels $1,\, 2,\, 3$, respectively.

In order to study the phase diagram of the system we make use of the fidelity $F$ and the fidelity susceptibility $\chi$ of neighboring states~\cite{gu10, SimInNat}, defined by
\begin{eqnarray}
	&&F(\tau,\,\tau+\delta\tau) = \vert\langle\psi(\tau)\vert\psi(\tau+\delta\tau)\rangle\vert^2\,, \nonumber \\[0.2in]
	&&\chi = 2\,\frac{1-F(\tau,\,\tau+\delta\tau)}{(\delta\tau)^2}\,.
	\label{fidelity}
\end{eqnarray}
Whereas the fidelity is a measure of the distance between states which vary as functions of a control parameter $\tau$, the fidelity susceptibility, essentially its second derivative with respect to the control parameter, is a more sensitive quantity. The fidelity measure goes to zero at each phase transition, as the nature of the ground state changes completely and orthogonaly; the fidelity susceptibility has divergences at these critical points in phase space. Crossing a separatrix produces a change in the total excitation number $\langle M \rangle$.

To follow a similar procedure as for the $2$-level systems, and be able to study the phase diagram analytically, we consider as a variational trial state the direct product of Heisenberg-Weyl $HW(1)$ coherent states for the radiation part, $\vert \alpha \}=e^{\alpha\,{a}^{\dagger}}\,\vert 0\rangle$, and $U(3)$ coherent states constructed by taking the exponential of the lowering generators acting on the highest weight states of $U(3)$~\cite{cordero1}
\begin{equation}
	\vert\zeta\} := \big|[h_{1},h_{2},h_{3}]
        \gamma_{1},\gamma_{2},\gamma_{3}\big\}
       =e^{\gamma_{3} {A}_{21}}
	\, e^{\gamma_{2} {A}_{31}} \,  e^{\gamma_{1} {A}_{32}} 
         \vert\,[h_{1},h_{2},h_{3}] \rangle \ , \nonumber
	\label{gelfandstate}
\end{equation}
where $\vert\,[h_{1},\,h_{2},\,h_{3}]\rangle$ represents the highest weight state of the Gelfand-Tsetlin basis in the irreducible representation $[h_{1},h_{2},h_{3}]$ of $U(3)$~\cite{gelfand50}. When we consider the totally symmetric representation $[N_A,0,0]$ (for indistinguishable particles) the trial state becomes
\begin{equation}
	\vert\alpha;\,\zeta\rangle = \vert \alpha;N_{a},\gamma_{2},\gamma_{3}\}
	=e^{\alpha\,{a}^{\dagger}}\vert 0\rangle \otimes
	\,e^{\gamma_{3}{A}_{21}}
	\,e^{\gamma_{2}{A}_{31}}
    \vert [N_{a},0,0] \rangle
\nonumber
\end{equation}
where the parameter $\gamma_1$ no longer appears since ${A}_{32}\,\vert\,[N_{a},\,0,\,0] \rangle=0$.

It is convenient to use a polar form for the complex parameters
\begin{equation}
	\alpha:=\rho\,\exp(i\,\phi)\ ,\quad
	\gamma_{j}:=\rho_{j}\,\exp(i\,\phi_{j})\ ,\
	j=1,\,2,\,3\ ,
    \label{polarform}
\end{equation}
and minimising with respect to these new parameters the energy surface $\mathcal{H}(\alpha,\,\zeta) =\{\alpha;\,\zeta\vert\, H\, \vert\alpha;\,\zeta\}/
\{\alpha;\,\zeta\vert
\alpha;\,\zeta\}$ in the RWA approximation
takes the form
\begin{eqnarray}
	\fl
	&&\mathcal{H}_{\hbox{\tiny{RWA}}}(\rho_{c},\,\rho_{2c},\,\rho_{3c})=
	\frac{1}{N_A}\Omega\,\rho_{c}^{2}+\Big\{\,\Big[\omega_{1}
	+\omega_{2}\,\rho_{2c}^{2}+\omega_{3}\,\rho_{3c}^{2}
	\Big]\nonumber\\
	\fl
	&&\qquad -\frac{2}{\sqrt{N_A}}\,\rho\,\Big[\mu_{12}\,\rho_{2c}
	+\mu_{13}\,\rho_{3c}
	+\mu_{23}\,\rho_{2c}\,\rho_{3c}\Big]\Big\}/
	\left(1+\rho_{2c}^{2}+\rho_{3c}^{2}\right)\ ,
	\label{energysurface3RWA}
\end{eqnarray}
where $\rho_{c}$, $\rho_{2c}$ and $\rho_{3c}$ denote the
critical values of the corresponding variables, and we have taken $\rho_1=1$. It is important to stress that \eref{energysurface3RWA} is valid for {\it all} three configurations.

\begin{figure}
	\begin{center}
	\scalebox{0.35}{\includegraphics{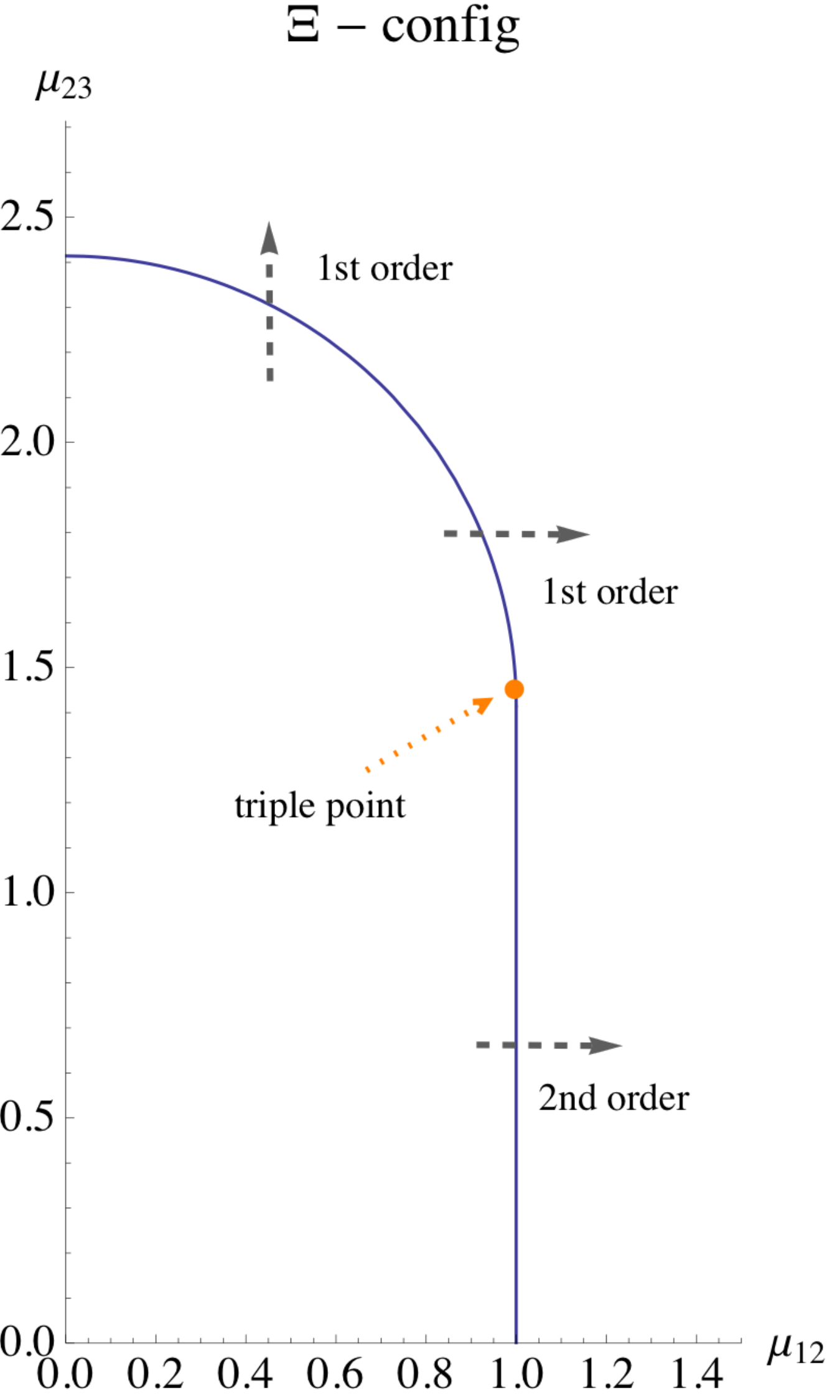}} \qquad\qquad
	\scalebox{0.35}{\includegraphics{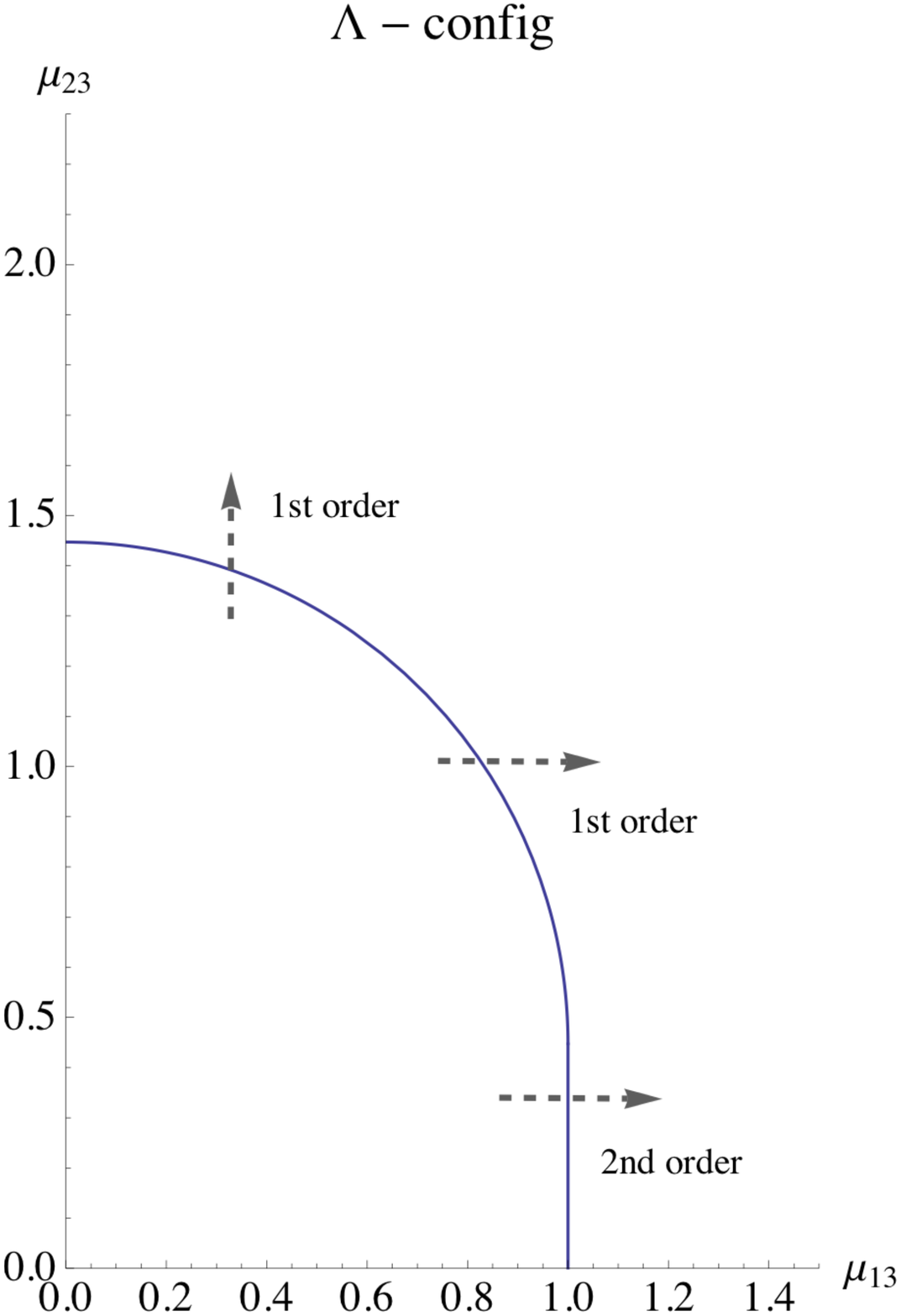}} \\
	\vspace{0.15in}
	\quad \scalebox{0.35}{\includegraphics{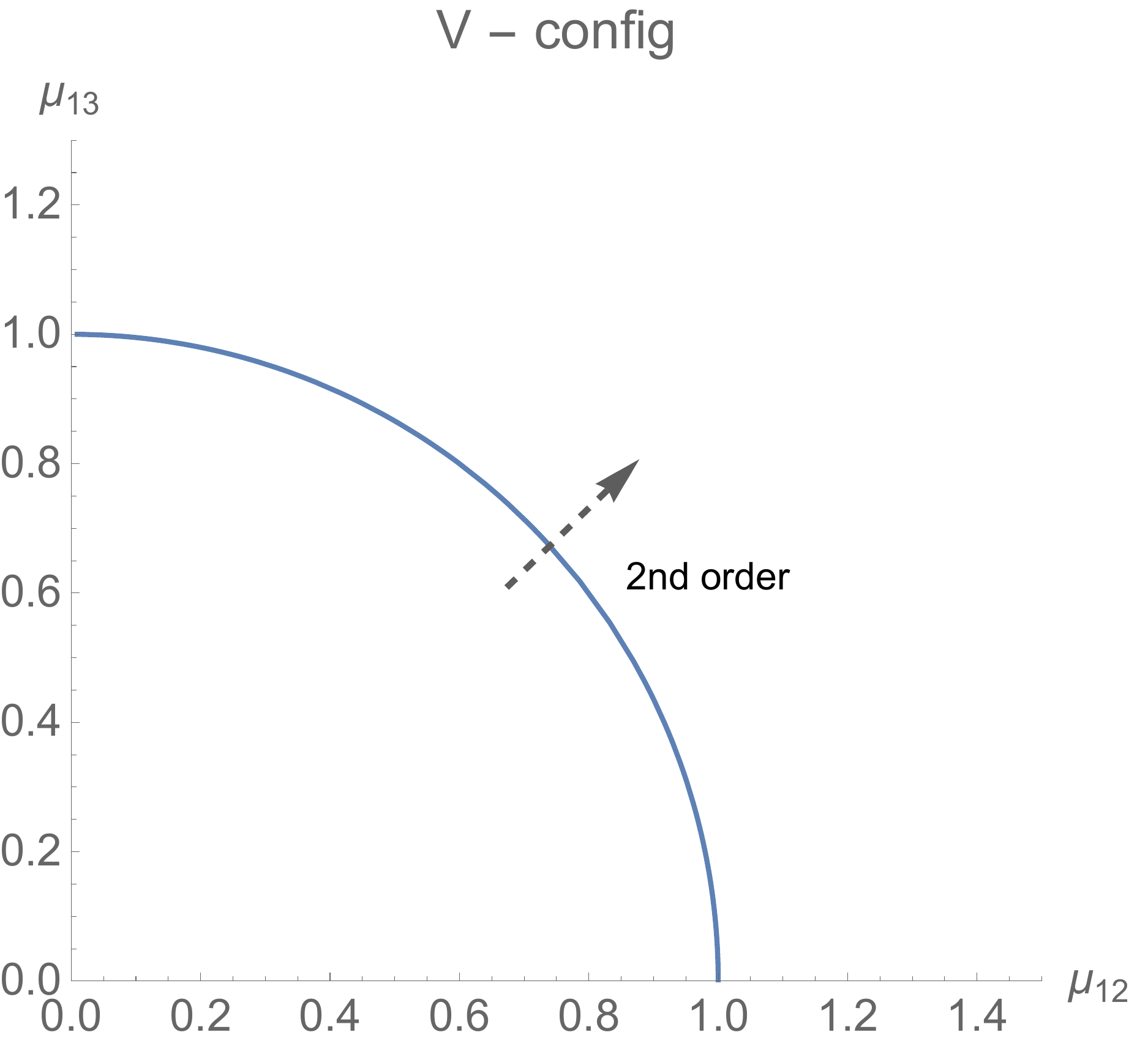}} \quad
	\scalebox{0.3}{\includegraphics{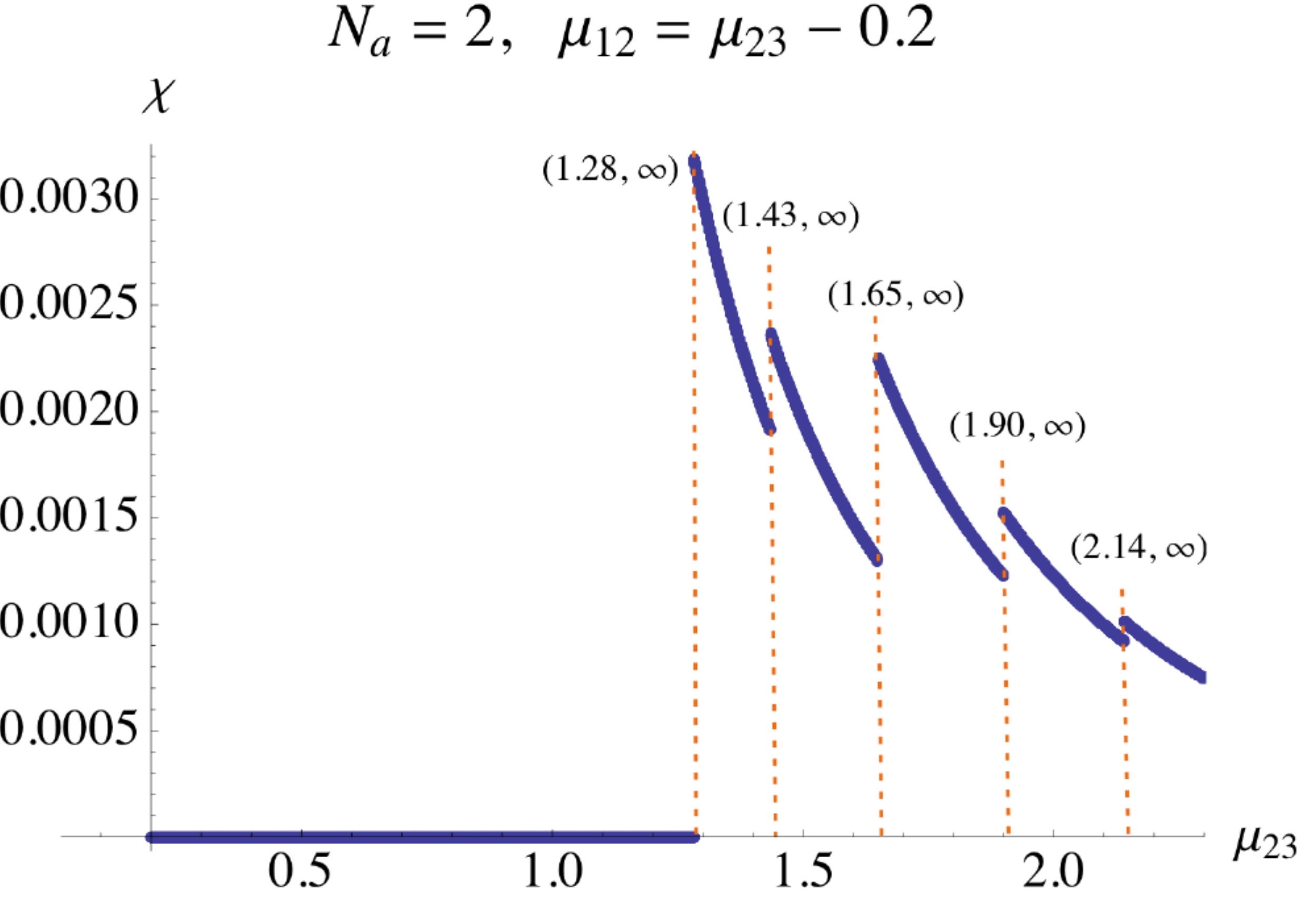}}
	\end{center}
	\caption{Shape of the separatrix $M=0 \to M\neq 0$ in phase space for the different configurations. We have taken double atomic resonance, except for the $\Lambda$-configuration where we took a detuning of $0.2$ between one of the transitions and the field frequency. The triple point in the $\Xi$-configuration is shown (see text). The fidelity susceptibility for neighbouring states (lower right) is shown for the (projected) SAS states in the $\Xi$-configuration along the path $\mu_{12}=\mu_{23}-0.2$.}
\label{separatrices3Levels}
\end{figure}

From this minimal surface the first separatrix corresponding to the phase change $M=0 \to M\neq 0$ (i.e., from the normal to the collective regimes) is given by~\cite{cordero1}
\begin{itemize}
	\item[i)]
for the $\Xi$--configuration
    \begin{equation}
	\mu_{12}^{2}+\left[\,\left|\mu_{23}\right|-\sqrt{\omega_{31}}\,
	\right]^{2}\,\Theta\left(\left|\mu_{23}\right|-\sqrt{\omega_{31}}
	\right)=\omega_{21}\ ;
	\label{sepXi}
    \end{equation}
	\item[ii)]
for the $\Lambda$--configuration
    \begin{equation}
	\mu_{13}^{2}+\left[\,\left|\mu_{23}\right|-\sqrt{\omega_{21}}\,
	\right]^{2}\,\Theta\left(\left|\mu_{23}\right|-\sqrt{\omega_{21}}
	\right)=\omega_{31}\ ;
	\label{sepL}
    \end{equation}
	\item[iii)]
for the $V$--configuration
    \begin{equation}
	\frac{\mu_{12}^{2}}{\omega_{21}}
	+\frac{\mu_{13}^{2}}{\omega_{31}}=1\ ;
	\label{sepV}
    \end{equation}
\end{itemize}
where $\omega_{ij} = \omega_i - \omega_j$ and $\Theta$ is the Heaviside function. These are shown in Fig.~\ref{separatrices3Levels} for double atomic resonance with respect to the radiation field frequency in the $\Xi$-configuration (i.e., $\omega_{31} = 2\,\omega_{21} = 2$), a small detuning in the $\Lambda$-configuration ($\omega_{21}=0.2,\ \omega_{31}=1$), and double atomic resonance in the $V$-configuration ($\omega_{21}=\omega_{31}=1$). For equal atomic detuning in the $\Lambda$-configuration the separatrix is identical to that of the $V$-configuration.

Further analysis shows that in the $\Xi$-configuration the phase transition across $\mu_{12}=\sqrt{\omega_{21}}$ is of second-order, while that across the segment of the circumference is of first-order. In the $\Lambda$-configuration with unequal atomic detuning we have the same behaviour. In the $\Lambda$-configuration with equal atomic detuning and in the $V$-configuration, however, all transitions are of second-order.

Being $\langle M \rangle = \nu + \lambda_2 \langle A_{22} \rangle + \lambda_3 \langle A_{33} \rangle$ a constant of motion, we obtain states adapted to the symmetry of the Hamiltonian by projecting onto the appropriate value of $\langle M \rangle$. This is done in practice by substituting $\nu = \langle M \rangle - \lambda_2 \langle A_{22} \rangle - \lambda_3 \langle A_{33} \rangle$ and keeping the only relevant value of $\langle M \rangle$. These are the (projected) SAS states in the RWA approximation.

\begin{figure}
	\begin{center}
	\scalebox{0.3}{\includegraphics{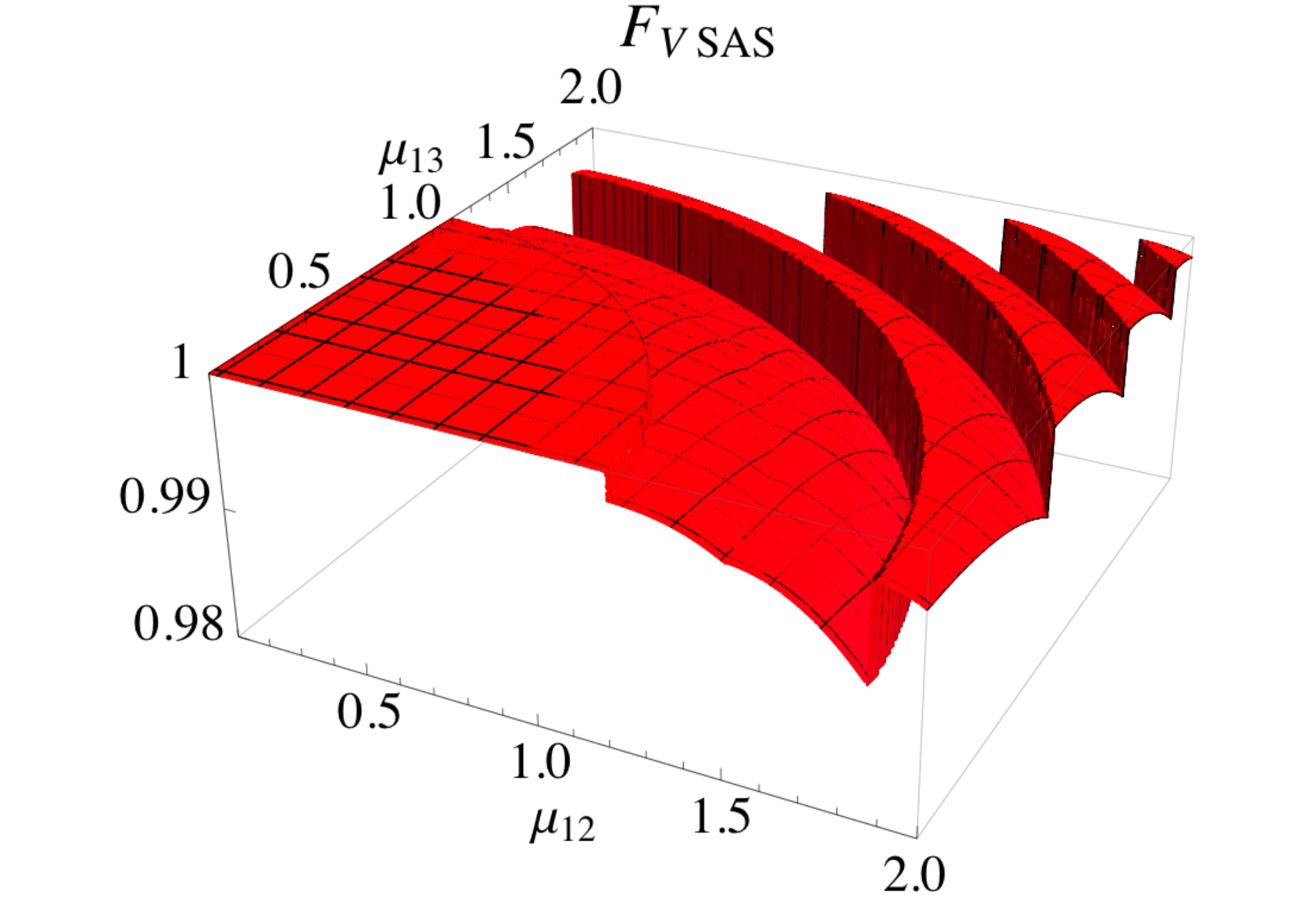}} \qquad\qquad
	\scalebox{0.3}{\includegraphics{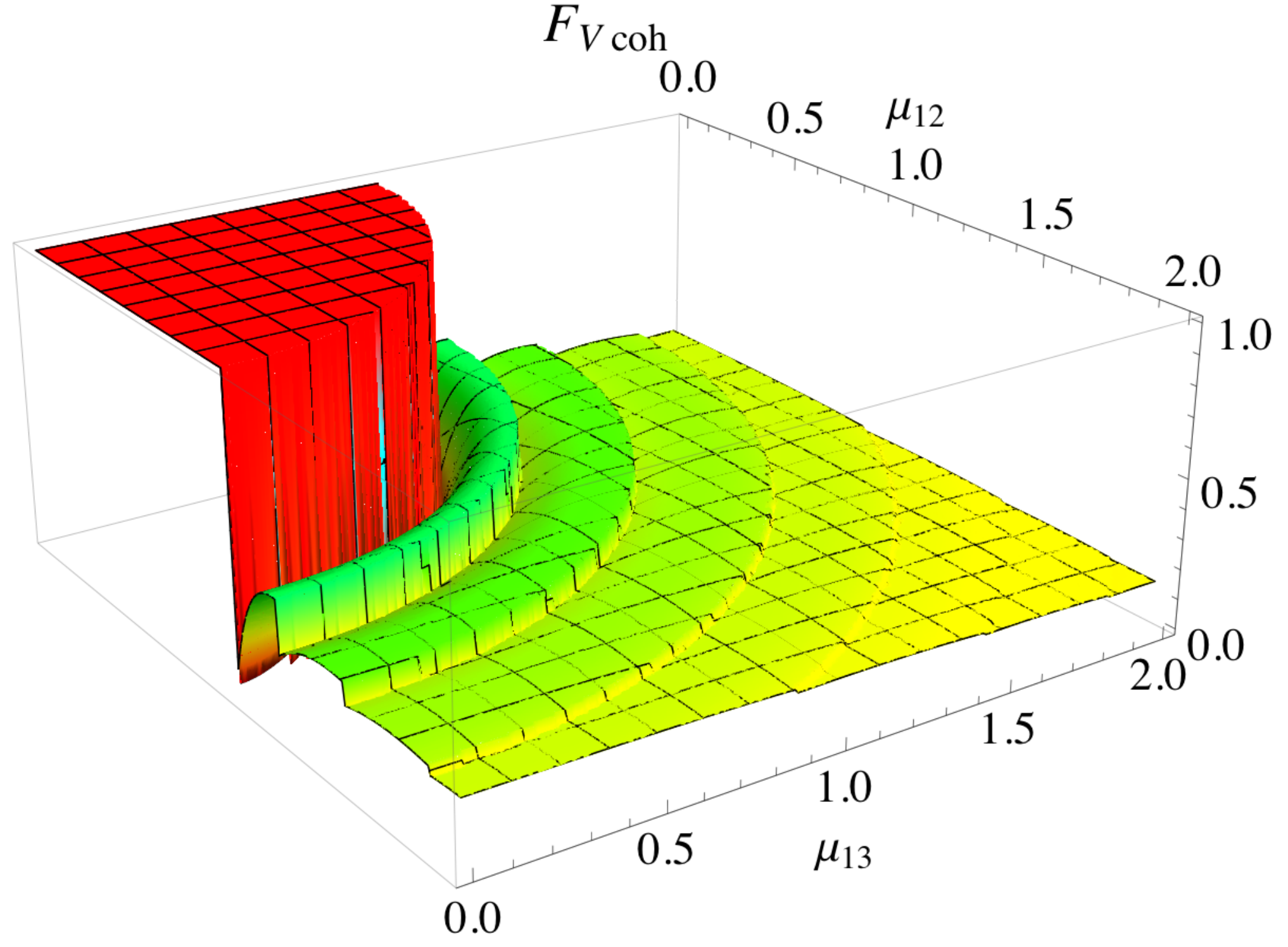}} \\
	\end{center}
	\caption{Fidelity between the quantum and (projected) SAS ground states (left), and that between the quantum and coherent ground states (right), for $N_A=3$, $V$-configuration.}
\label{fidelity3levels}
\end{figure}

In the thermodynamic limit, given by $\nu \propto N_A$ with $N_A\to\infty$, the loci in parameter space of the {\it quantum} phase transitions are exactly those shown by Fig.~\ref{separatrices3Levels}. But even for a small number of atoms the approximation to the separatrices given by the projected SAS states is remarkably good: the figure shows, in its lower right, the fidelity susceptibility divergences at each phase crossing along the path $\mu_{12}=\mu_{23}-0.2$ for the $\Xi$-configuration and $N_A=2$, as a function of $\mu_{23}$. Since $\mu_{12}=1$ is fixed and independent of $N_A$ at this separatrix ({\it vide infra}), the projected SAS prediction gives $\mu_{23}=1.2$ which compares well with the value of $\mu_{23}=1.28$ for the first transition of the exact quantum ground state, even though $N_A=2$. This good approximation by the chosen variational states obeys the fact that the fidelity between the quantum and projected SAS ground states gives a perfect overlap except in a small vicinity of the phase transitions, as shown in Figure~\ref{fidelity3levels} (left). The reader may compare this vs. the overlap between the quantum and the coherent ground states shown at right. In both cases $N_A=3$ and we have chosen to illustrate the result for the $V$-configuration (those for the other configurations being very similar).

\begin{figure}
	\begin{center}
	\scalebox{0.28}{\includegraphics{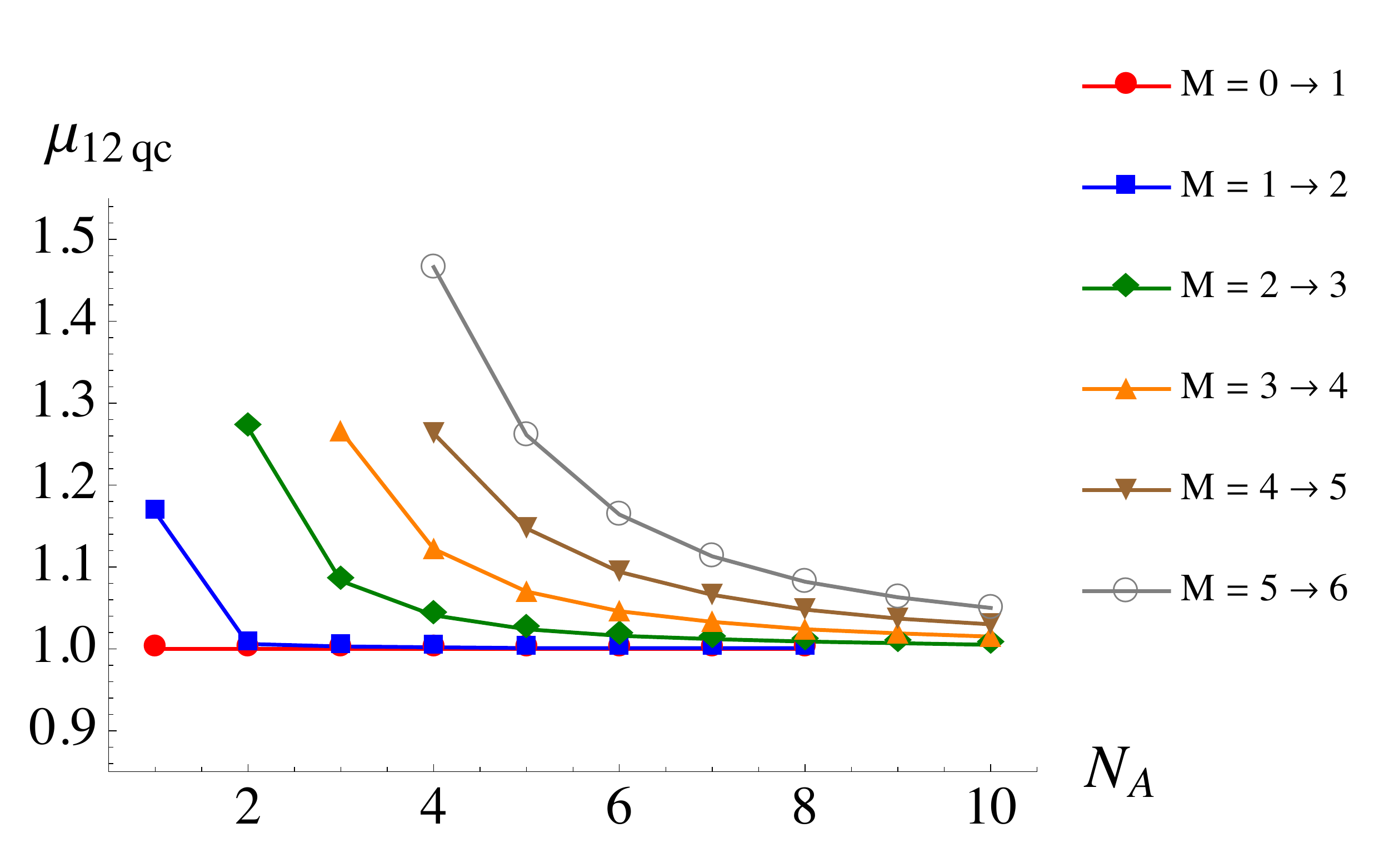}} \qquad
	\scalebox{0.28}{\includegraphics{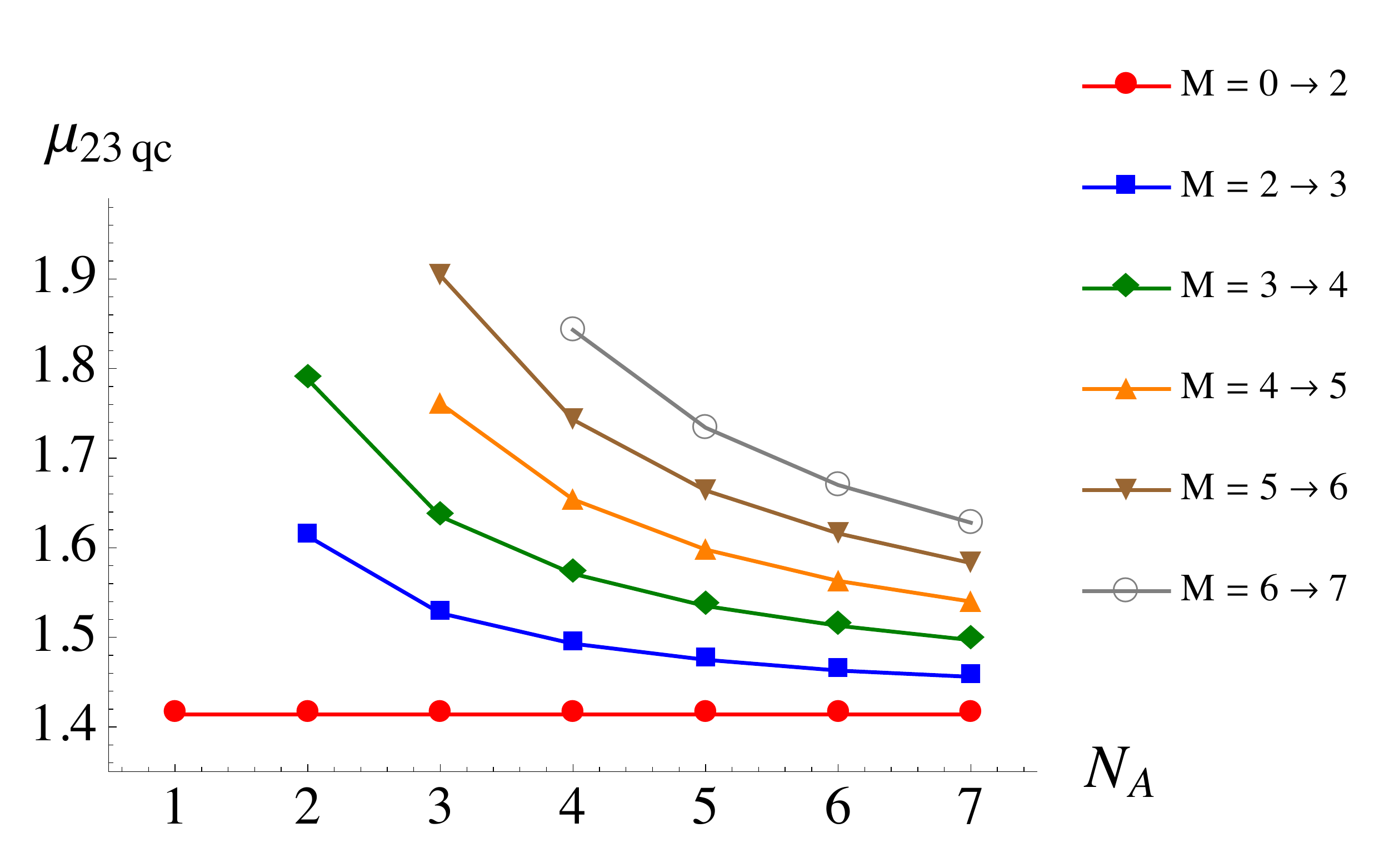}}
	\end{center}
	\caption{Critical values $\mu_{ij\,qc}$ of the interaction parameters for the quantum transitions $M \to M+1$ as functions of the number of atoms in the $\Xi$-configuration. This shows how the loci of the quantum phase transitions change as the number of atoms grow. In the limit $N_A\to\infty$ they converge to the separatrix between the normal and collective regions.}
\label{TransitionsMu_vs_Na}
\end{figure}

All phase transitions tend to those given by equations~(\ref{sepXi}, \ref{sepL}, \ref{sepV}) as $N_A\to\infty$ with $\nu \propto N_A$. In this thermodynamic limit these are the only ones that remain. For the $V$-configuration the consequent transitions take place at a family of curves congruent with and ever more distant to the one shown in Fig.~\ref{separatrices3Levels}, which approach the latter uniformly as $N_A$ grows. For the $\Xi$-configuration we have curves with similar shape to that shown in Fig.~\ref{separatrices3Levels}, with a vertical straight edge and an upper circular arc. The vertical edge tends to that at $\mu_{12}=1$ as $N_A$ grows, while the circular arcs ``slide down'' the $\mu_{23}$-axis, intersect the arc of the transition $M=0 \to M\neq 0$ shown, and continues sliding down this arc tending to $(\mu_{12},\,\mu_{23})=(1,\,\sqrt{2})$ as $N_A\to\infty$. We show this in Figure~\ref{TransitionsMu_vs_Na}. The subfigure at left shows the critical value $\mu_{12\,qc}$ of $\mu_{12}$ for the quantum transition $M \to M+1$ as a function of the number of atoms, i.e., how the transitions to the right of the straight vertical line $\mu_{12}=1$ move as $N_A$ changes. They all tend to the limit $\mu_{12\,qc}=1$, as given by \eref{sepXi} when $N_A\to\infty$. At right we plot $\mu_{23\,qc}$ as a function of $N_A$, to see how the phase transitions above the circular arc move; the first transition is $M=0 \to M=2$ since the phase region $M=1$ stops at $\mu_{23}=\sqrt{2}$ and does not reach the upper arc. We see that the point where these phase regions intersect the circular arc slide towards $\mu_{23}=\sqrt{2}$, again as given by \eref{sepXi}. In the thermodynamic limit, then, the separatrix reduces to the line segment given by $\mu_{12}=1$ and $\mu_{23} \in [0,\,\sqrt{2}]$, plus the arc of circumference starting at $\mu_{23}=\sqrt{2}$.

The $\Lambda$-configuration has a similar behaviour as the $V$-configuration when in double resonance, and a behaviour much like that of the $\Xi$-configuration when away from double resonance.

\subsection{A Triple Point in Phase Space}

\begin{figure}
	\begin{center}
	\scalebox{0.45}{\includegraphics{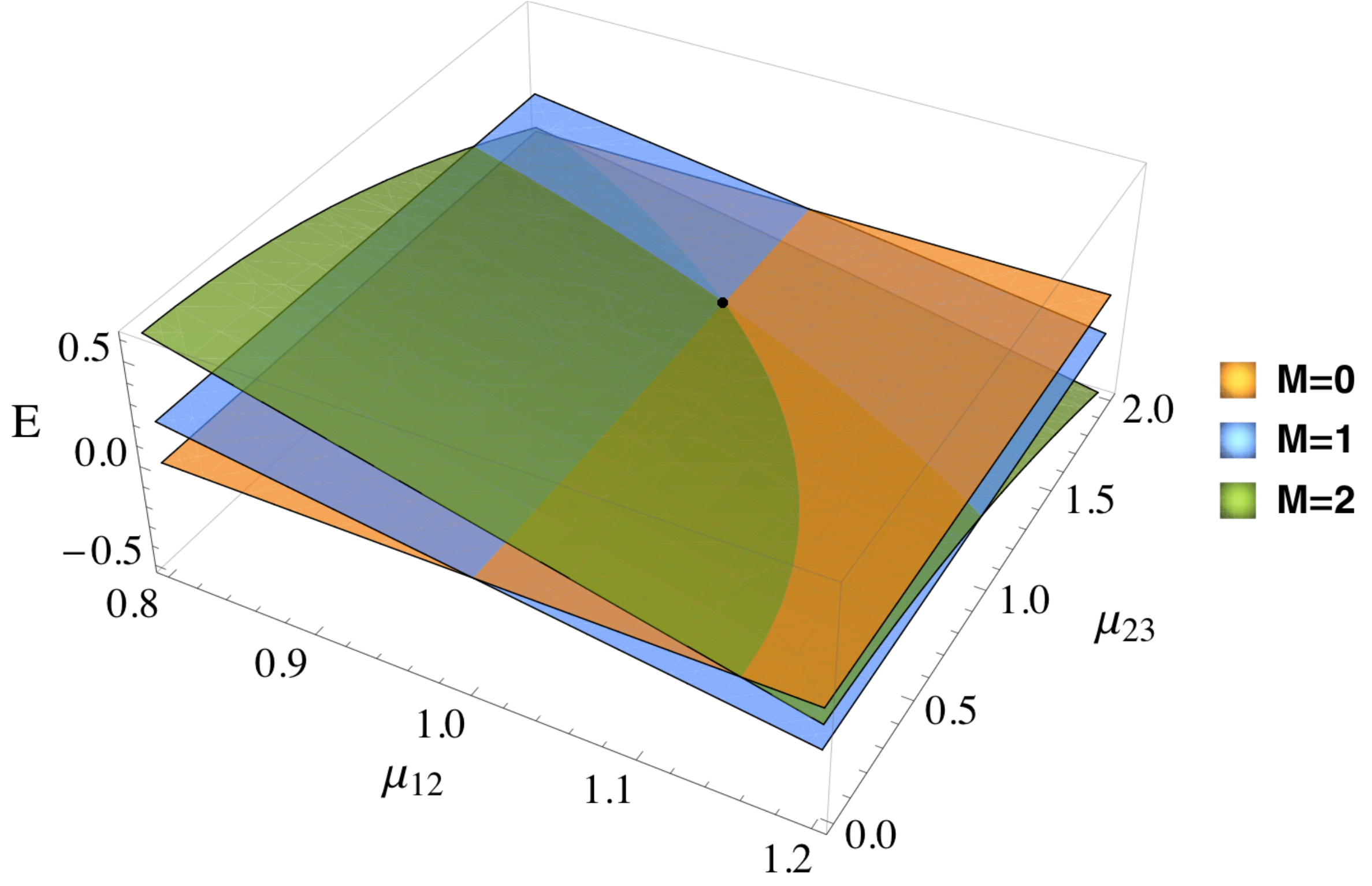}}
	\end{center}
	\caption{Energy of the ground state plotted as a function of $\mu_{12}$ and $\mu_{23}$ for the $\Xi$-configuration in double resonance. The 3 regions meet at a point, the {\it triple point}, at coordinates $(1,\sqrt{2},\,0)$ in parameter space (marked in the figure with a black dot). This point is independent of the number of atoms, and subsists in the thermodynamic limit.}
\label{TriplePointPlot}
\end{figure}

The $\Xi$-configuration is special in that it shows a richer structure. In particular, it has a {\it triple point} in parameter space, corresponding to the place where the phases for $M=0$, $M=1$, and $M=2$ meet~\cite{TriplePoint}. The term {\it triple point} is mainly used in the context of fluids, where different phases of the fluid meet in parameter space. Here we use the same terminology since the different values of the total excitation number $M$ correspond to completely different structures of the ground state, even though the energy is the same for all of them, and since these three regions meet at a point in parameter space as shown in Figure~\ref{TriplePointPlot}. The meaning is also the same as for a thermodynamic triple point: any fluctuation (in this case quantum) will drastically change the composition of the ground state. And since in the collective region we have a decay rate proportional to $N_A{}^2$, as opposed to $N_A$ for the normal region, this gives hope for experimental exploitation of the triple point.

In the RWA approximation and in double resonance this triple point resides at $(\mu_{12},\,\mu_{23}) = (1,\sqrt{2})$ (cf. Fig.~\ref{separatrices3Levels}). In the full model, contemplating the counter-rotating terms, we just divide these values by $2$ ({\it vide infra, Subsection~\ref{FullModel}}). It is a {\it fixed} point, independent of $N_A$, which subsists in the thermodynamic limit. It is also characteristic of the $\Xi$-configuration; it does not appear in the $\Lambda$ or the $V$ configurations. We can calculate the ground state $\vert\psi\rangle_{gs}$ at the triple point for each phase, by diagonalising the Hamiltonian in the basis $\vert\nu;\,q,\,r\rangle$. In the RWA approximation one gets analytic expressions. For $N_A\geq 2$ we have:
\numparts
\begin{itemize}
	\item[]
		M=0\,:
		\begin{equation}
		\fl
		\vert\psi\rangle_{gs} = \vert 0;\,N_A,\,N_A\rangle
		\end{equation}
		\vspace{0.1in}
	\item[]
		M=1\,:
		\begin{equation}
		\fl
		\vert\psi\rangle_{gs} = \frac{1}{\sqrt{2}}\,\vert 0;\,N_A,\,N_A-1\rangle + \frac{1}{\sqrt{2}}\,\vert 1;\,N_A,\,N_A\rangle
		\end{equation}
		\vspace{0.1in}	
	\item[]
		M=2\,:
		\begin{eqnarray}
		\fl
		\vert\psi\rangle_{gs} = && -\frac{1}{2\sqrt{N_A}}\,\vert 0;\,N_A-1,\,N_A-1\rangle + \frac{1}{2}\sqrt{\frac{N_A-1}{N_A}}\,\vert 0;\,N_A,\,N_A-2\rangle + \nonumber\\
		\fl
		&& + \frac{1}{\sqrt{2}}\,\vert 1;\,N_A,\,N_A-1\rangle + \frac{1}{2}\,\vert 2;\,N_A,\,N_A\rangle
		\end{eqnarray}
\end{itemize}
\endnumparts
It is clear that, even when we are at the same point in phase space, the ground state may acquire very different structures. Away from double resonance the triple point is still present, though its coordinates in phase space vary as well as the specific combination given by the equations above.

\begin{figure}
	\begin{center}
	\scalebox{0.45}{\includegraphics{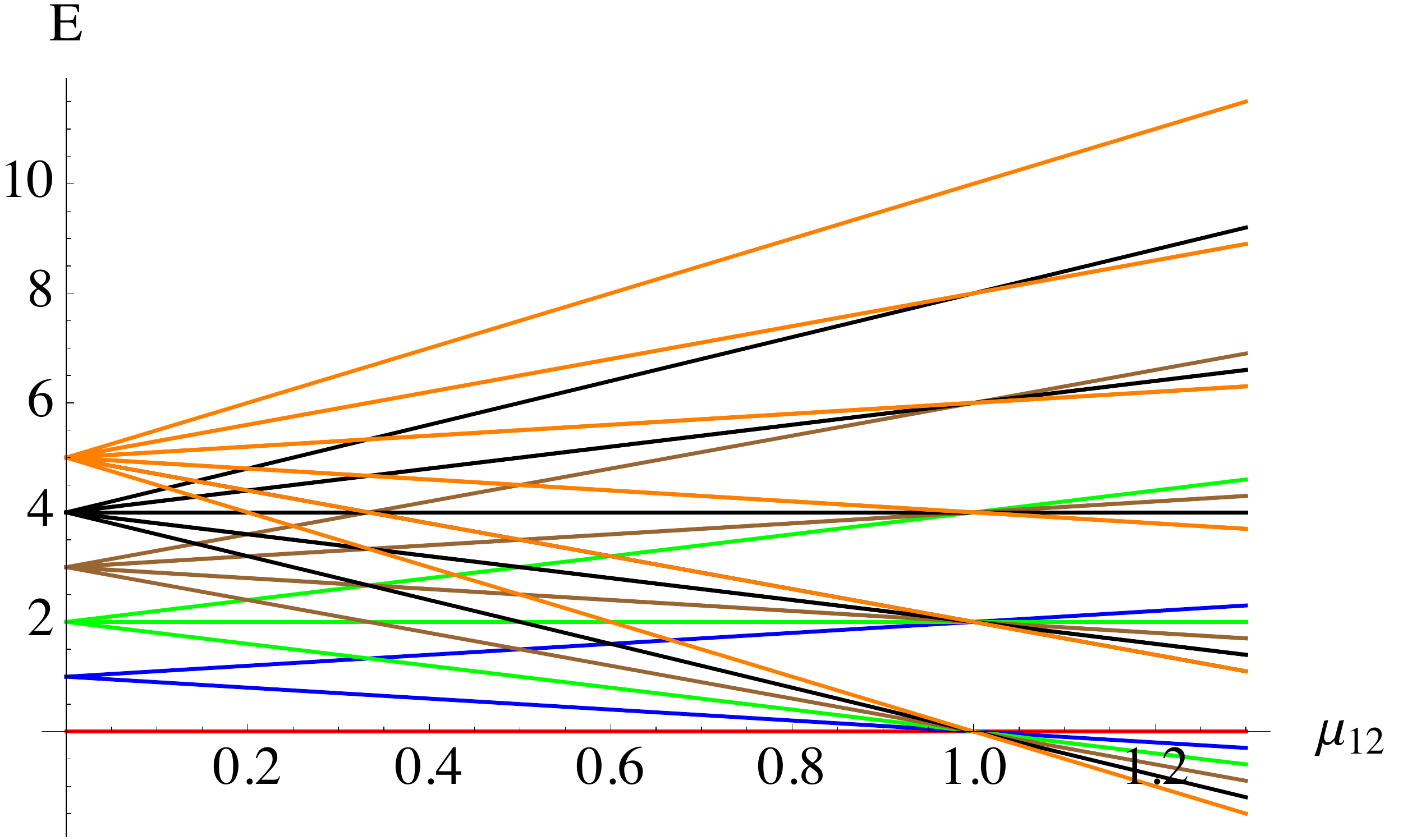}}
	\end{center}
	\caption{Colour online. Energy spectrum in the limit $N_A\to\infty$ of the $\Xi$-configuration. A collapse of energy levels for all values of $M$ at precisely $\mu_{12}=1$, the surviving separatrix in phase space, is clear. Different colours correspond to different values of $M$; equally, the value of $M$ can be read as the value for the energy at $\mu_{12}=0$.}
\label{spectrumPT}
\end{figure}

When the number of excitations $M$ is small the dimension of the Hilbert space does not depend on $N_A$, making it possible to study the system in the limit $N_A\to\infty$. The energy spectrum, in particular, does not depend on $\mu_{23}$ in this limit, and it shows a collapse of energy levels at precisely $\mu_{12}=1$ for all values of $M$. Figure~\ref{spectrumPT} shows this for $M=0$ to $M=5$, and it is interesting to compare it with the spectrum of the $2$-level Tavis-Cummings model, Fig.~\ref{spectrumTCM}. As a function of $M$, at the triple point, we have an equidistant spectrum with only even harmonics~\cite{TriplePoint}, and it is interesting to note that at $\mu_{12}=1$ we have precisely all the even harmonics as degenerate energy levels, and no others.

\begin{figure}
	\begin{center}
	\scalebox{0.28}{\includegraphics{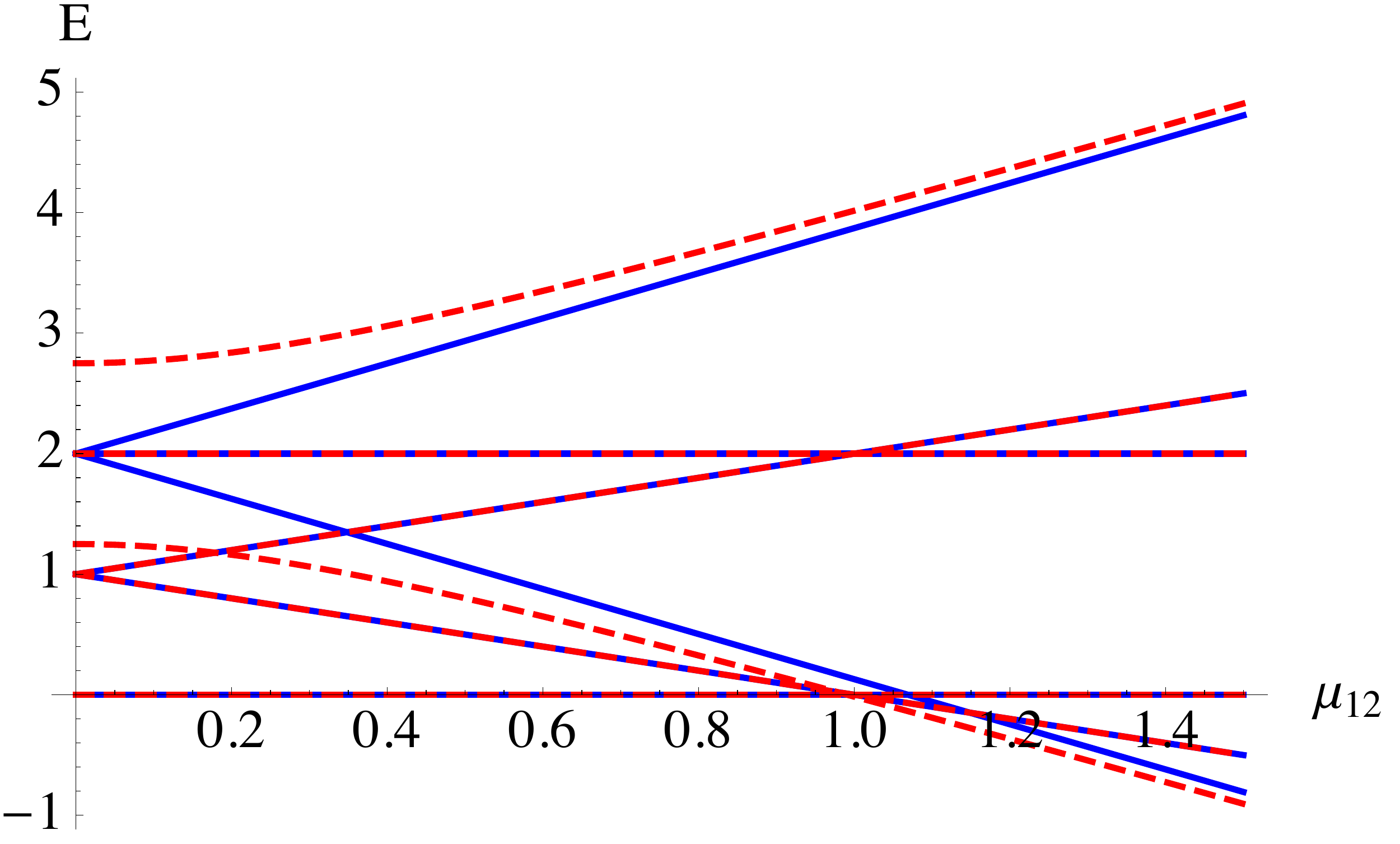}} \quad
	\scalebox{0.28}{\includegraphics{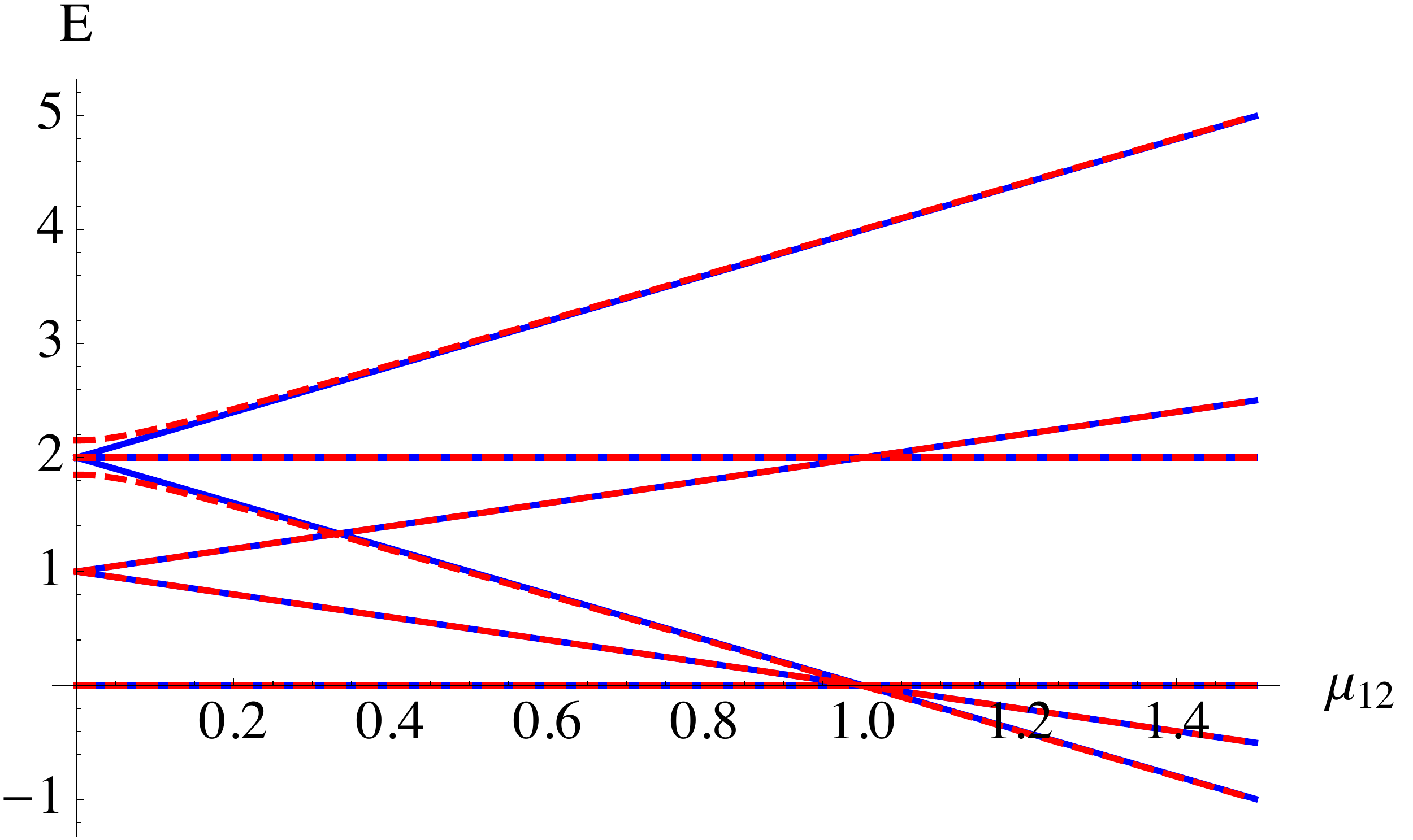}}
	\end{center}
	\caption{Colour online. Energy spectrum degeneracies at $\mu_{12}=0$ broken when $N_A$ is finite. Shown are $N_A=4$ (left) and $N_A=100$ (right), when $\mu_{23}=0$ (blue, continuous line) and when $\mu_{23}=1.5$ (red, dashed line).}
\label{degBreak}
\end{figure}

The behaviour at the axis $\mu_{12}=0$ is also interesting. The total degeneracy for each $M$ found at $\mu_{12}=0$ in the limit $N_A\to\infty$ only survives for finite $N_A$ at $\mu_{23}=0$, i.e., when there is absolutely no matter-field coupling. As soon as the coupling is ``turned on'', this degeneracy breaks down. Figure~\ref{degBreak} shows $E$ vs. $\mu_{12}$ for $N_A=4$ (left) and $N_A=100$ (right), when $\mu_{23}=0$ (blue, continuous line) and when $\mu_{23}=1.5$ (red, dashed line). While the ground and first excited states still show degeneracies at $\mu_{12}=0$, these are broken for the second excited state.

\subsection{Counter-Rotating Terms: the full model}
\label{FullModel}

When we do {\it not} make the rotating wave approximation, i.e., we include the counter-rotating terms in equations (\ref{Hamiltonian3Levels}, \ref{Hint}), minimising the energy surface $\mathcal{H}(\alpha,\,\zeta) =\{\alpha;\,\zeta\vert\, H\, \vert\alpha;\,\zeta\}/
\{\alpha;\,\zeta\vert
\alpha;\,\zeta\}$ with respect to the polar parameters takes the form~\cite{Ramon}
\begin{eqnarray}
	\fl
	&&\mathcal{H}(\rho_{c},\,\rho_{2c},\,\rho_{3c})
	=\frac{1}{N_A}\Omega\,\rho_{c}^{2}+\Big\{\,\Big[\omega_{1}
	+\omega_{2}\,\rho_{2c}^{2}+\omega_{3}\,\rho_{3c}^{2}
	\Big]\nonumber\\
	\fl
	&&\qquad \qquad \quad -\frac{4}{\sqrt{N_A}}\, \,\rho_{c}\Big[\mu_{12}\,\rho_{2c}
	+\mu_{13}\,\rho_{3c}
	+\mu_{23}\,\rho_{2c}\,\rho_{3c}\Big]\Big\}/
	\left(1+\rho_{2c}^{2}+\rho_{3c}^{2}\right)\, ,
	\label{energysurface3}
\end{eqnarray}

Comparing equations (\ref{energysurface3}, \ref{energysurface3RWA}), the energy surfaces $\mathcal{H}$ and $\mathcal{H}_{\hbox{\tiny{RWA}}}$ coincide if we identify
\begin{equation}
	\left(\mu_{jk}\right)_{\hbox{\tiny{RWA}}}
	\longrightarrow 2\,\left(\mu_{jk}\right) \ .
	\label{muidentification}
\end{equation}
This means that $\mathcal{H}_{\hbox{\tiny{RWA}}}$ will inherit the properties of $\mathcal{H}$ at values of $(\mu_{ij})_{\hbox{\tiny{RWA}}}$ equal to
$\frac{1}{2}\mu_{ij}$. (This is the same behaviour as that mentioned earlier for the Dicke model.) In particular, the shape of the phase diagram will be inherited in full at coordinates half those of the RWA scenario, and the order of the phase transitions will be the same.

Whereas $M$ is a constant of motion in the RWA approximation, it is not in the full model. As in the $2$-level DM model, it is the {\it parity} in the number $\langle M \rangle$ of excitations that is conserved, as $U(\theta):=\exp\left(i\,\theta\,M\right)$ is only a symmetry operator for $\theta=0,\ \pi$. To obtain {symmetry-adapted} states we therefore take linear combinations of coherent states of the same parity
\begin{equation}
	\vert\alpha;\,\zeta\}_{\pm}:=\left(\mathbf{1}
	\pm\exp[i\,\pi\,M]\right)
	\,\vert\alpha;\,\zeta\}\ ,
\end{equation}
and the energy surface for these SAS states, in {\it any} configuration, results in~\cite{Ramon}
\begin{eqnarray}
	\fl
\mathcal{H}_{\pm}&=&{}_{\pm}\{\alpha;\,\zeta\vert\, H \,\vert\alpha;\,\zeta\}_{\pm}
   \nonumber\\
   \fl
   &=&\frac{2}{N_A}\,\Omega\,|\alpha|^{2}\,\Big[\exp(|\alpha|^{2})\, (\gamma^{\ast}\cdot\gamma)^{N_A}
  \mp\exp(-|\alpha|^{2})\,(\gamma^{\ast}\cdot\tilde{\gamma})^{N_{A}}\Big]\nonumber\\
   \fl
   &&+2\,\sum_{i=1}^{3}\omega_{i}\,|\gamma_{i}|^{2}\,\Big[
   \exp(|\alpha|^{2})\,(\gamma^{\ast}\cdot\gamma)^{N_A-1}
   \pm(-1)^{\lambda_{i}}\,\exp(-|\alpha|^{2})
   \,(\gamma^{\ast}\cdot\tilde{\gamma})^{N_A-1}\Big]\nonumber\\
   \fl
   &&+\frac{1}{\sqrt{N_A}}\,(\alpha+\alpha^{\ast})\,\sum_{i<j=1}^{3}\,\mu_{ij}
   \,(1-(-1)^{\lambda_{i}+\lambda_{j}})\nonumber\\
   \fl
   &&\quad\times\Big[\exp(|\alpha|^{2})\,(\gamma_{i}^{\ast}\,\gamma_{j}
   +\gamma_{j}^{\ast}\,\gamma_{i})\,(\gamma^{\ast}\cdot\gamma)^{N_A-1}
   \nonumber\\
   \fl
  &&\qquad\pm\exp(-|\alpha|^{2})\,((-1)^{\lambda_{i}}\,\gamma_{i}^{\ast}\,\gamma_{j}
   +(-1)^{\lambda_{j}}\,\gamma_{j}^{\ast}\,\gamma_{i})
   \,(\gamma^{\ast}\cdot\tilde{\gamma})^{N_A-1}\Big]\ ,
	\label{energysurface3SAS}
\end{eqnarray}
where $\gamma = (\gamma_1,\,\gamma_2,\,\gamma_3),\ \tilde\gamma = (\gamma_1,\,(-1)^{\lambda_2} \gamma_2,\,(-1)^{\lambda_3} \gamma_3)$, and $\gamma_1=1$. Again, one may use the polar form of these parameters to minimise with respect to each one in order to obtain the minimum energy surface for the system. In general this has to be done numerically, but the $V$-configuration lends itself to an analytic treatment; furthermore, all transitions in this configuration are of second order, making it a good candidate for the study of its critical exponents.

\subsection{Critical Exponents}

\begin{figure}
	\begin{center}
	\scalebox{0.26}{\includegraphics{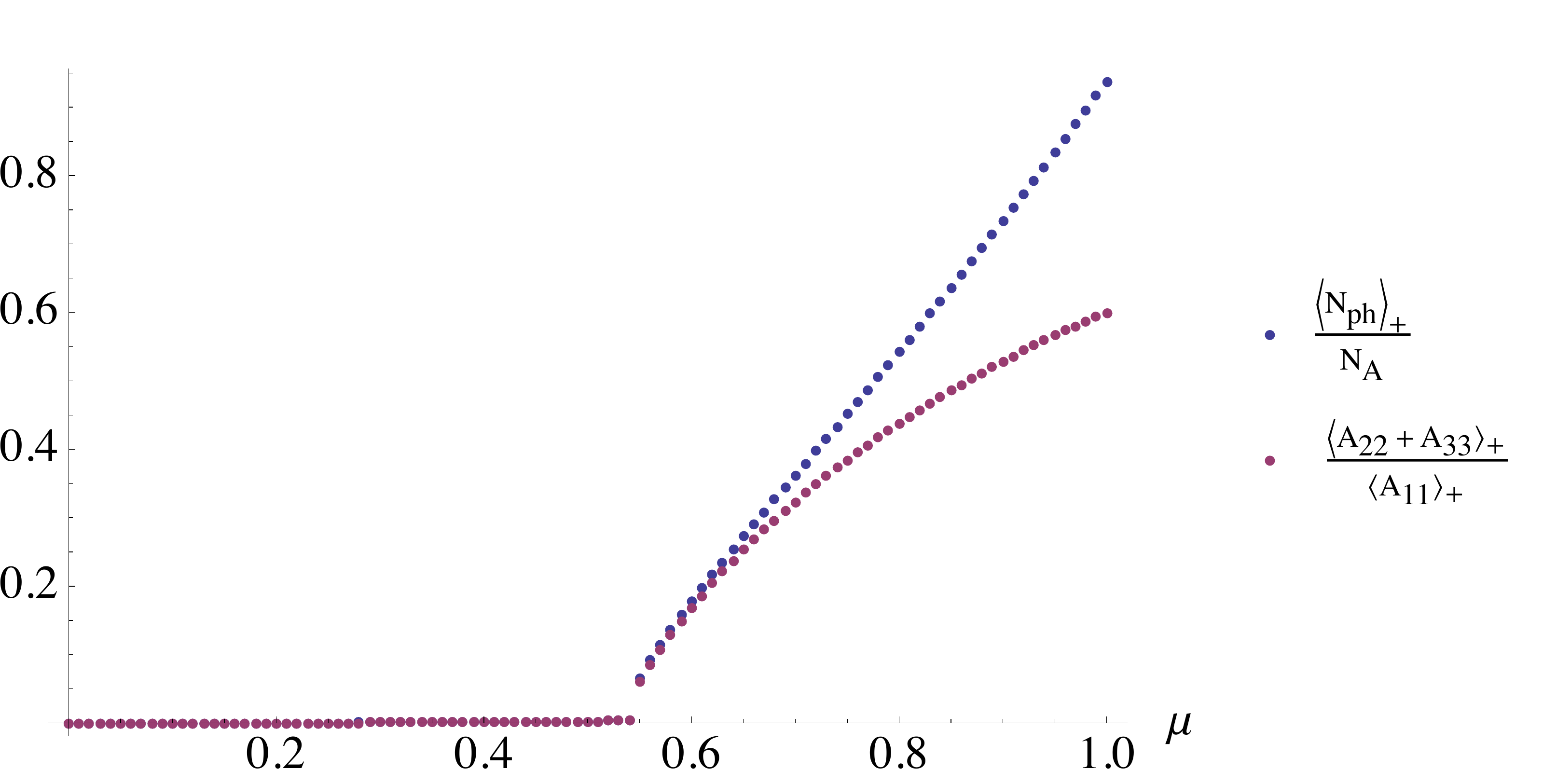}} \quad
	\scalebox{0.26}{\includegraphics{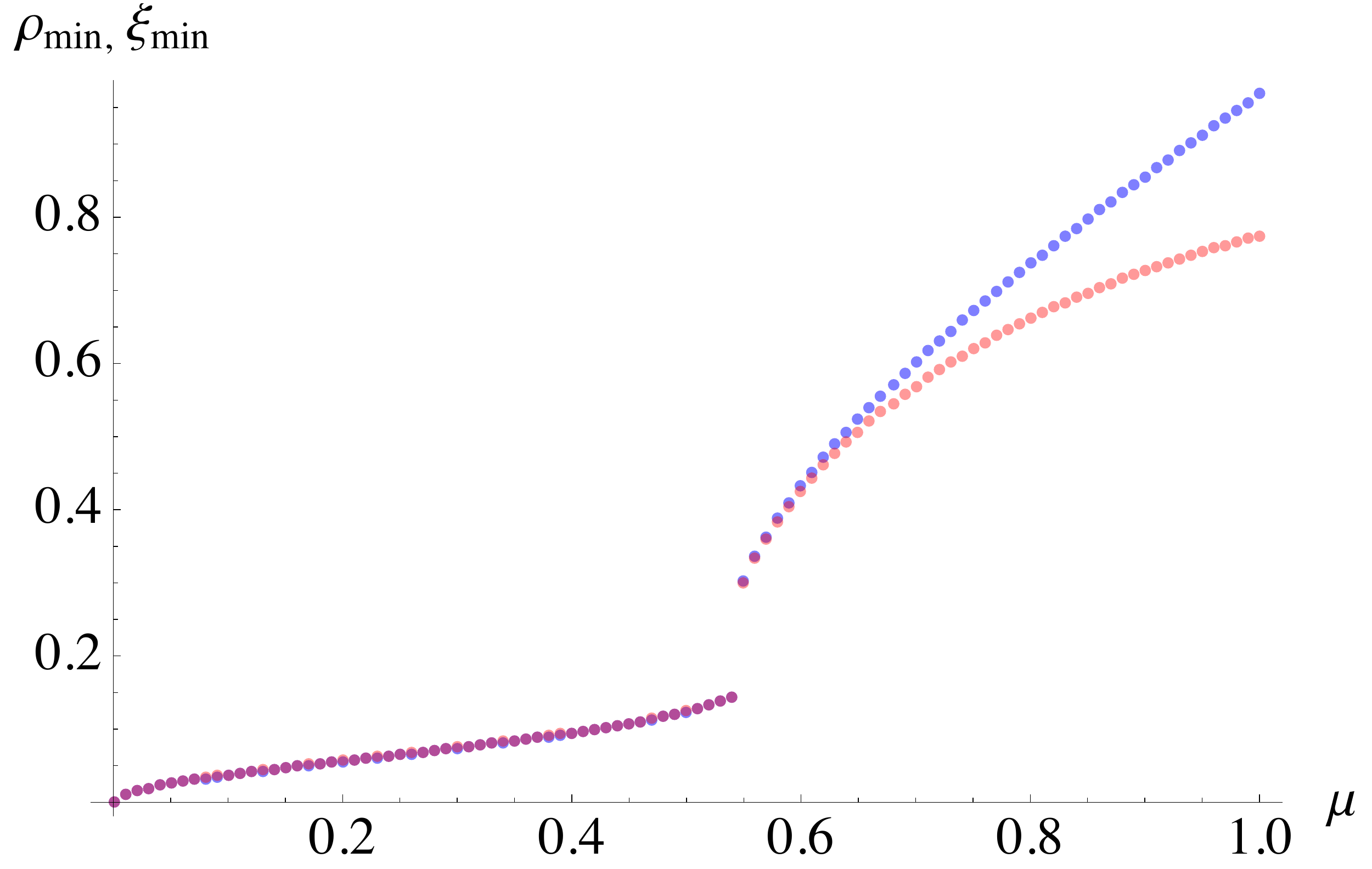}}
	\end{center}
	\caption{Colour online. Left: equality of the (normalised) atomic and field excitations in the normal regime; upper (blue) curve corresponds to the field excitations and lower (red) curve to the atomic excitations. Right: discontinuity in the phase parameters at a phase transition for the symmetry-adapted ground state in the $V$-configuration; upper (blue) curve corresponds to $\rho_{\hbox{min}}$ and lower (red) curve to $\xi_{\hbox{min}}$.}
\label{PTV}
\end{figure}

Using the polar form given in \eref{polarform}, and further defining
\begin{eqnarray}
	\rho_2 &=& \xi\,\cos(\eta), \qquad \rho_3=\xi\,\sin(\eta),\nonumber\\
	\mu_{12} &=& \mu\,\cos(\theta), \qquad \mu_{13}=\mu\,\sin(\theta),\nonumber\\
	\chi &=& \eta - \theta
\end{eqnarray}
the SAS energy surface (\ref{energysurface3SAS}) in the $V$-configuration takes the form
\begin{eqnarray}
	\fl
	\mathcal{H}_{V\,+} =&& \frac{1}{\mathcal{K}}
	\left[ -\left(1-\xi^2\right)^{N_A} \left(1+\xi^2\right) \left(-\xi^2+\rho^2\left(-1+\xi^2\right)\right) \right. \nonumber\\
	\fl
	 &&\left. + \e^{2N_A\rho^2} \left(-1+\xi^2\right)\left(1+\xi^2\right)^{N_A} \left(\xi^2+\rho^2\left(1+\xi^2\right)-4\,\rho\,\xi\,\mu\,\cos(\chi)\right) \right]
\end{eqnarray}
where we have defined
\begin{equation}
	\mathcal{K} = \left[ \left(-1+\xi^4\right) \left( \left(1-\xi^2\right)^{N_A} + \e^{2N_A\rho^2} \left(1+\xi^2\right)^{N_A} \right) \right]\ .
\end{equation}
We minimise with respect to the parameters $(\rho,\,\xi)$, and having an analytical expression for the SAS ground state allows to evaluate the relevant field and atomic operators in this state. Of particular interest is the comparative behaviour of the system in the normal and collective regimes: Figure~\ref{PTV} (left) shows the (normalised) atomic number of excitations in comparison with the (normalised) field excitations. This suggests that, in the normal region,
\begin{equation}
\frac{\langle A_{22} + A_{33} \rangle_+}{\langle A_{11} \rangle_+} =
\frac{\langle N_{ph} \rangle_+}{N_A}
\label{DecayRate}	
\end{equation}
as is to be expected from the atomic decay rate in a non-collective regime; as soon as the system enters a collective regime the number of field excitations increases much more rapidly than the atomic excitations. Differently to the coherent states, the SAS and quantum ground states in the normal regime contain non-zero contributions of photonic and atomic excitations. \Eref{DecayRate} thus may be used, in general, as a criterion to {\it define} the normal region by using the appropriate atomic operators for the atomic excitations in each configuration.

\begin{figure}
	\begin{center}
	\scalebox{0.45}{\includegraphics{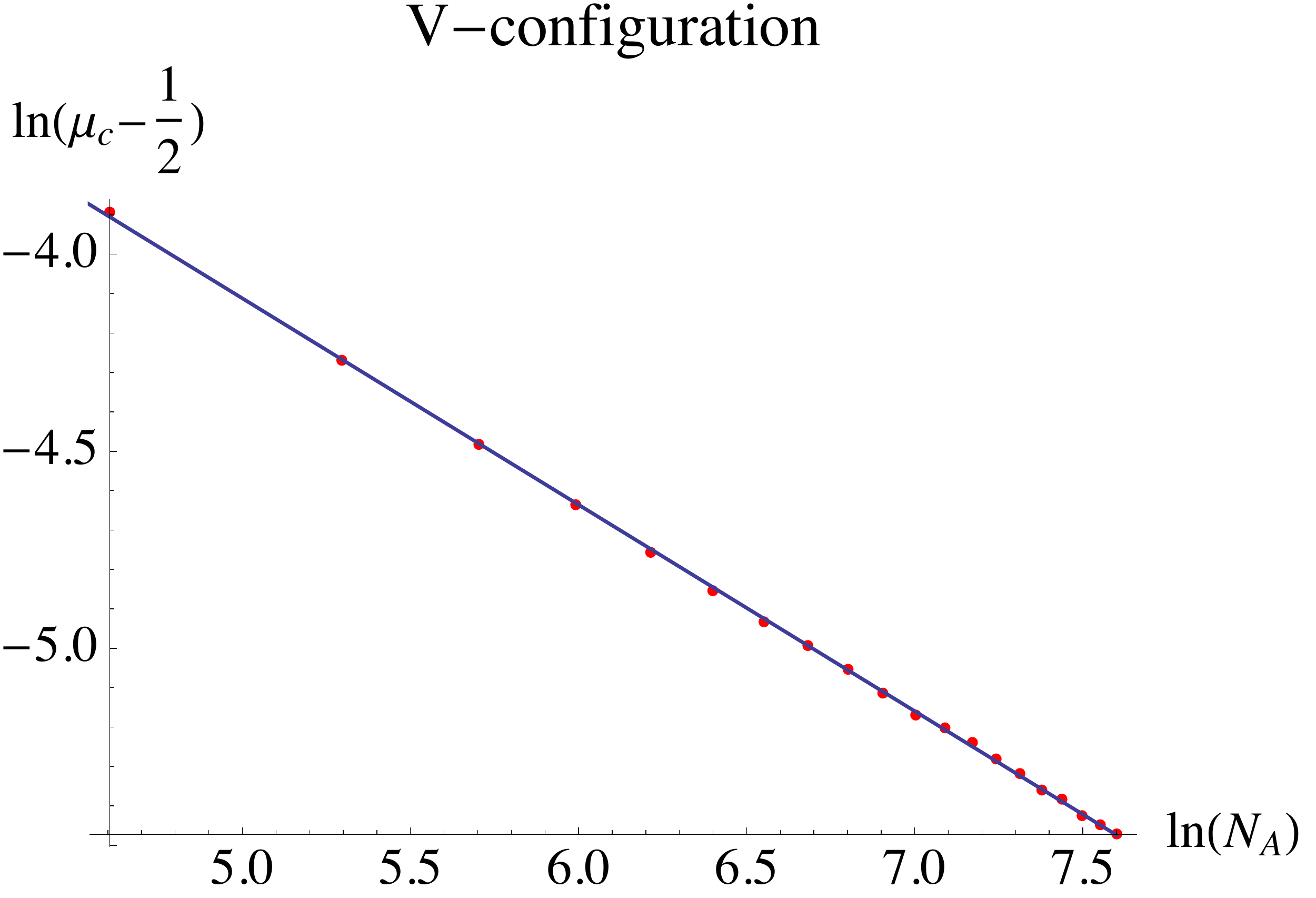}}
	\end{center}
	\caption{Logarithmic behaviour of the critical value of the coupling parameter $\mu_c$ with the number of atoms $N_A$, for the symmetry-adapted ground state in the $V$-configuration. The linear fit (continuous straight line) shows a critical exponent of $-11/21$, exactly as that found for the Dicke model using SAS states.}
\label{logmu_vs_logNa_3_sas}
\end{figure}

A phase transition causes a change in the structure of the ground state, which is reflected by a discontinuity in the phase parameters (see Figure~\ref{PTV} (right)). 
We use this discontinuity to find the critical value of the interaction strength $\mu_c$ at the phase transition that separates the normal from the collective regimes as a function of the number of atoms, from $N_A=100$ to $N_A=2000$. Figure~\ref{logmu_vs_logNa_3_sas} shows a logarithmic plot for these two variables, together with the linear fit
\begin{equation}
	\mu_c^{sas} = \frac{1}{2} + \e^{-\frac{3}{2}}\, N_A^{-\frac{11}{21}}\, .
	\label{ExpCritSAS_3}
\end{equation}
%

It is interesting to compare this relation with that obtained for the Dicke model using SAS states, \eref{ExpCritSAS}. The critical exponent is exactly the same. An analysis of residuals shows a confidence interval of $[-0.530,\,-0.519]$ for the exponent, to a confidence level of $0.95$, where $-\frac{11}{21}$ fits perfectly, with a goodness-of-fit $R^2 = 0.9997$.

\section{Discussion and Conclusions}

We have reviewed and expanded upon the structure of the phase diagram for systems consisting of $2$- and $3$-level particles dipolarly interacting with a $1$-mode electromagnetic field, inside a cavity, paying particular attention to the case of a {\it finite} number $N_A$ of particles, and showing that the divergences that appear in other treatments are a consequence of the mathematical approximations employed, and can be avoided by studying the system in an exact manner quantum-mechanically or via a catastrophe formalism with variational trial states that satisfy the symmetries of the appropriate Hamiltonians.

We have shown how the use of these variational states give an excellent approximation not only to the exact quantum phase space, but also to the energy spectrum and the expectation values of the atomic and field operators. Furthermore, they allow for {\it analytic} expressions in many of the cases studied, even for finite $N_A$. We have made use of the fidelity and the fidelity susceptibility of neighbouring quantum states to find the loci of the transitions in phase space from one phase to the other; having analytic expressions allows for the order of the quantum phase transitions to be determined explicitly for each of the configurations, with and without detuning. Finally, we have derived the critical exponents for the various systems.

The $\Xi$-configuration in $3$-level systems is particular in that it exhibits a {\it triple point} in phase space. This means that any quantum fluctuation at this location will drastically change the composition of the ground state. The exact form of the ground state at this triple point has been studied; the same can be done for excited states in the vicinity of this point or elsewhere in parameter space.

Finally, a criterion (\eref{DecayRate}) to define the normal region for the full Hamiltonian in the different configurations was suggested, acknowledging the fact that the SAS and quantum ground states in the normal regime contain non-zero contributions of photonic and atomic excitations.

With the promise of the benefits of quantum information the study of these systems acquire greater importance, as they constitute the basic q-dit blocks themselves as well as the possible quantum logical gates for computational purposes. The properties of the systems treated here have been intriguing, no less because of the search for a fine control of the light-matter interaction at the level of  single and few atom-photon pairings. It is hoped that this manuscript conveys an accurate account of the properties and structure of the phase space of these interesting systems.

\section*{Acknowledgments}
This work was partially supported by CONACyT-M\'exico (under project 238494), and by DGAPA-UNAM (under projects IN101614 and IN110114).


\vspace{0.2in}

\appendix

\section{Expectation values and fluctuations: Dicke Model}
\label{appA}

The following table shows the expectation values and fluctuations of matter and field observables for the coherent and symmetry-adapted states in
the collective (superradiant) regime. The mean-field behaviour for the
normal region can be recovered by taking the limit $x\rightarrow 1$. (Adapted from~\cite{erratum}.)

Comparing \eref{energysurface} and \eref{energysurfaceDicke} it is clear that, by substituting $\gamma_c \to \gamma_c/2$ in the Table, we obtain the expectation values of operators for the Tavis-Cummings model, and their fluctuations.

\begin{landscape}
	\begin{table}[h!]
	\caption{\label{tab1}
	Expectation values and fluctuations of matter and field
	observables for the coherent and symmetry-adapted states in
	the superradiant regime. The mean-field behaviour obtained in the
	normal region can be recovered by taking the limit
	$x\rightarrow 1$.}
	\vspace{0.15in}
	\centering
	\resizebox{0.8\columnwidth}{!}{
	\begin{tabular}{c|c|c}
		\raisebox{-1.5ex}[0pt]
		&Coherent
		&Symmetry Adapted\\
		\hline
		\hline
		\rule[-3mm]{0mm}{5mm}
		$\langle{q}\rangle$
		&$-\sqrt{2N_A}\,\gamma_{c}\,x\,\sqrt{1-x^{-4}}$
		&$0$\\
		\rule[-3mm]{0mm}{5mm}
		$\langle{p}\rangle$
		&$0$
		&$0$\\
		\rule[-3mm]{0mm}{5mm}
		$\langle {J}_{x}\rangle$
		&$\frac{N_A}{2}\,\sqrt{1-x^{-4}}$
		&$0$\\
		\rule[-3mm]{0mm}{5mm}
		$\langle {J}_{y}\rangle$
		&$0$
		&$0$\\
		\rule[-3mm]{0mm}{5mm}
		$\langle {J}_{z}\rangle$
		&$-\frac{N_A}{2}\,x^{-2}$
		&$-\frac{N_A}{2}\,x^{2}\left(1-\frac{1-x^{-4}}{1\pm \mathcal{F}}
		\right)$\\
		\rule[-3mm]{0mm}{5mm}
		$\langle{a}^{\dagger}{a}\rangle$
		&$N_A\,\gamma_{c}^{2}\,x^{2}\,\left(1-x^{-4}\right)$
		&$N_A\,\gamma_{c}^{2}\,x^{2}\,\left(1-x^{-4}\right)\left(\frac{
		1\mp \mathcal{F}}{1\pm \mathcal{F}}\right)$\\
		\rule[-3mm]{0mm}{5mm}
		$\langle\hat{\Lambda}\rangle$
		&$\frac{N_A}{2}\left(1-x^{-2}+2\,\gamma_{c}^{2}\,x^{2}\,
                        \left(1-x^{-4}\right)\right)$
		&\phantom{x}$\frac {N_A} {2}\left(\frac {1 - x^{-2}} {1\pm \mathcal{F}} \right)
		\left[1 + 2 \gamma_ {c}^{2} \left (1 + x^{2} \right)
     		\mp\left (x^{2} + 
      		2\gamma_ {c}^{2}\left (1 + x^{2} \right) \right)
                 \mathcal{F} \right]$\phantom{x}\\
		\hline
		\rule[-3mm]{0mm}{5mm}
		$(\Delta{q})^{2}$
		&$\frac{1}{2}$
		&$\frac{1}{2}+2N_A\,\gamma_{c}^{2}\,x^{2}\,\left(\frac{
                        1-x^{-4}}{1\pm \mathcal{F}}\right)$\\
		\rule[-3mm]{0mm}{5mm}
		$(\Delta{p})^{2}$
		&$\frac{1}{2}$
		&$\frac{1}{2}\mp 2N_A\,\gamma_{c}^{2}\,x^{2}\,\left(
                        \frac{1-x^{-4}}{1\pm \mathcal{F}}\right)\,\mathcal{F}$\\
		\rule[-3mm]{0mm}{5mm}
		$(\Delta{J}_{x})^{2}$
		&$\frac{N_A}{4}\,x^{-4}$
		&$\frac{N_A}{4}\left(1+\frac{\left(N_A-1\right)\left(
		1-x^{-4}\right)}{1\pm \mathcal{F}}\right)$\\
		\rule[-3mm]{0mm}{5mm}
		$(\Delta{J}_{y})^{2}$
		&$\frac{N_A}{4}$
		&$\frac{N_A}{4}\left(1\pm\frac{\left(N_A-1\right)
		\left(1-x^{4}\right)\mathcal{F}}{1\pm \mathcal{F}}\right)$\\
		\rule[-3mm]{0mm}{5mm}
		$(\Delta{J}_{z})^{2}$
		&$\frac{N_A}{4}\,\left(1-x^{-4}\right)$
		&$\frac{N_A}{4}\,\frac{\left(1-x^{-4}\right)}{\left(1\pm
		\mathcal{F}\right)^{2}}\,\left[1\mp\left(N_A-1\right)\left(1
		-x^{4}\right)\,\mathcal{F}-x^{4}\mathcal{F}^{2}\right]$
		\\
		\rule[-3mm]{0mm}{5mm}
		$(\Delta\,{a}^{\dagger}{a})^{2}$
		&$N_A\,\gamma_{c}^{2}\,x^{2}\,\left(1-x^{-4}\right)$
		&$\frac{N_A\,\gamma_{c}^{2}\,x^{2}\,\left(1-x^{-4}\right)}{1\pm \mathcal{F}}
		\left[1\mp \mathcal{F}\pm 4\,N_A\,\gamma_{c}^{2}
                 \,x^{2}\,\left(1-x^{-4}\right)\frac{\mathcal{F}}{1\pm \mathcal{F}}\right]$
                 \\
		\hline
		\rule[-3mm]{0mm}{5mm}
		$\langle {J}_{z}\,{a}^{\dagger}{a}\rangle$
		&$-\frac{N_A^2}{2}\,\gamma_{c}^{2}\,\left(1-x^{-4}\right)$
		&$-\frac{N_A^2}{2}\,\gamma_{c}^{2}\,x^{4}\left(1-x^{-4}\right)
                        \left(\frac{x^{-4}\mp \mathcal{F}}{1\pm \mathcal{F}}\right)$\\
		\rule[-3mm]{0mm}{5mm}
		$\langle {J}_{x}\,{q}\rangle$
		&$-\sqrt{\frac{N_A^{3}}{2}}\,\gamma_{c}\,x\,
		\left(1-x^{-4}\right)$
		&$-\sqrt{\frac{N_A^{3}}{2}}\,\gamma_{c}\,x\,
                        \frac{1-x^{-4}}{1\pm \mathcal{F}}$\\
		\rule[-3mm]{0mm}{5mm}
		$(\Delta\,\hat{\Lambda})^{2}$
		&$\frac{N_A \left(1-x^{-4}\right)}{4}\, \left(1 + 4 \, \gamma_c^2 \, x^2 \right)
		$ 
		&$\frac{N_A\,\left(1-x^{-4}\right)}{4\,\left(1\pm \mathcal{F}\right)^{2}}\bigg[
		1+4\,x^2\,\gamma_c^2\pm \mathcal{F}\,\left(1-x^4\right) \left(1-N_A\,\left(1+4\,\gamma_c^2\right)^2 \right) $ \\
		& &$-x^2  \, \mathcal{F}^2 \left(x^2+4\,\gamma_c^2\right)\bigg]$\\
	\end{tabular}
	}
	\end{table}
\end{landscape}


\vspace{0.2in}


\section*{References}
\begin{thebibliography}{99}

\bibitem{cummings65}
Cummings F W 1965 {\it Phys. Rev.} {\bf 140} A1051

\bibitem{eberly80}
Eberly J H, Narozhny N B and S\'anchez-Mondrag\'on J J 1980 {\it Phys. Rev. Lett.} {\bf 44} 1323
	
\bibitem{rempe87}
Rempe G, Walther H and Klein N 1987 {\it Phys. Rev. Lett.} {\bf 58} 353

\bibitem{romera09}
Romera E and de los Santos F 2009 {\it Phys. Rev.} B {\bf 80} 165416

\bibitem{romera07}
Romera E and de los Santos F 2007 {\it Phys. Rev. Lett.} {\bf 99} 263601; 2008 {\it Phys. Rev.} A {\bf 78} 013837

\bibitem{robinett04}
Robinett R W 2004 {\it Phys. Rep.} {\bf 392} 1

\bibitem{jaynes63}
Jaynes E T and Cummings F W 1963 {\it Proc. IEEE} {\bf 51} 89
 
\bibitem{tavis69}
Tavis M and Cummings F W 1969 {\it Phys. Rev.} {\bf 170} 379; {\bf 188}, 692

\bibitem{dicke54}
Dicke R H 1954 {\it Phys. Rev.} {\bf 93} 99

\bibitem{camop2}
Nahmad-Achar E, {Casta\~nos} O, {L\'opez-Pe\~na} R and Hirsch J~G 2013 {\it
Phys. Scr.} {\bf 87} 038114

\bibitem{rzazewski75} 
Rzazewski K, Wodkiewicz K and Zakowicz W 1975 {\it Phys. Rev. Lett.} {\bf 35} 432

\bibitem{knight78} 
Knight J M, Aharonov Y and Hsieh G T C 1978 {\it Phys. Rev.} A {\bf 17} 1454; Bialynicki-Birula I and Rzazewski K 1979 {\it Phys. Rev.} A {\bf 19} 301; Gawedzki K and Rzazewski K 1981 {\it Phys. Rev. A} {\bf 23} 2134

\bibitem{domokos12}
Vukics A and Domokos P 2012 {\it Phys. Rev.} A {\bf 86} 053807

\bibitem{baumann10}
Baumann K, Guerlin C, Brennecke F and Esslinger T 2010 {\it Nature} {\bf 464} 1301

\bibitem{nagy10}
Nagy D, Konya G, Szirmai G and Domokos P 2010 {\it Phys. Rev. Lett.} {\bf104} 130401

\bibitem{hepp73}
Hepp K and Lieb E H 1973 {\it Ann. Phys.} {\bf 76} 360

\bibitem{sachdev}
Sachdev S 2011 {\sl Quantum Phase Transitions}, (Cambridge Univ. Press, Cambridge)

\bibitem{reslen05}
Reslen J, Quiroga L and Johnson N F 2005, {\it Europhys. Lett.} {\bf 69}, 8

\bibitem{zanardi06} 
Zanardi P and Paunkovi\'c N 2006 {\it Phys. Rev.} E {\bf 74}, 031123

\bibitem{goto08}
Goto H and Ichimura K 2008 {\it Phys. Rev.} A {\bf 77}, 053811

\bibitem{emary03}
Emary C and Brandes T 2003 {\it Phys. Rev.} E {\bf 67} 066203; {\it Phys. Rev. Lett.} {\bf 90} 044101

\bibitem{romera12a}
Romera E, {del Real} R and Calixto M 2012 {\it Phys. Rev.} A {\bf 85} 053831

\bibitem{romera12b}
Romera E, Calixto M and Nagy A 2012 {\it Europhys. Lett.} {\bf 97} 20011

\bibitem{gilmore81} 
Gilmore R and Narducci L 1978 {\it Phys. Rev.} A {\bf 17}, 1747; Gilmore R 1981 {\sl Catastrophe Theory for Scientists and Engineers}, (Wiley, New York)

\bibitem{scrip}
Casta\~nos O, L\'opez-Pe\~na R, Nahmad-Achar E, Hirsch J G, L\'opez-Moreno E and Vitela J E 2009 {\it Phys. Scr.} {\bf 79} 065405
 
\bibitem{scrip2}
Casta\~nos O, Nahmad-Achar E, L\'opez-Pe\~na R and Hirsch J G 2009 {\it Phys. Scr.} {\bf 80} 055401

\bibitem{papercorto}
Casta\~nos O, Nahmad-Achar E, L\'opez-Pe\~na R and Hirsch J G 2011 {\it Phys. Rev.} A {\bf 83} 051601(R)

\bibitem{paperextenso}
Casta\~nos O, Nahmad-Achar E, L\'opez-Pe\~na R and Hirsch J G 2011 {\it Phys. Rev.} A {\bf 84} 013819

\bibitem{universal}
Casta\~nos O, Nahmad-Achar E, L\'opez-Pe\~na R and Hirsch J G 2012 {\it Phys. Rev.} A {\bf 86} 023814

\bibitem{camop1}
Hirsch J G, Casta\~nos O, L\'opez-Pe\~na R and Nahmad-Achar E 2013 {\it
Phys. Scr.} {\bf 87} 038106

\bibitem{crossovers}
Hirsch J G, Casta\~nos O, Nahmad-Achar E and L\'opez-Pe\~na R 2013 {\it
Phys. Scr.} {\bf T153} 014033

\bibitem{gu10}
Gu S-J 2010 {\it Int. J. Mod. Phys.} B {\bf 24} 4371

\bibitem{cordero1}
Cordero S, L\'opez-Pe\~na R, Casta\~nos O and Nahmad-Achar E 2013 {\it
Phys. Rev.} A {\bf 87} 023805

\bibitem{baksic}
Baksic A, Nataf P and Ciuti C 2013 {\it Phys. Rev.} A {\bf 87} 023813

\bibitem{yi03}
Yi X~X, Su X~H and You L 2003 {\it Phys. Rev. Lett.} {\bf 90}(9) 097902

\bibitem{jane02}
Jan\'e E, Plenio M~B and Jonathan D 2002 {\it Phys. Rev.} A {\bf 65}(5) 050302

\bibitem{LAOP}
Casta\~nos O, Cordero S, L\'opez-Pe\~na R and Nahmad-Achar E 2014 {\it Latin America Optics and Photonics Conference (LAOP)} OSA Technical Digest (published online: http://www.opticsinfobase.org/search2.cfm?reissue=J\&journalList=\&fullrecord=casta\~nos\&basicsearch=Go)

\bibitem{manko74}
Dodonov V V, Malkin I A and Man'ko V I 1974 {\it Physica} {\bf 72} 597

\bibitem{manko01}
Man'ko M A, Man'ko V I and Mendes R V 2001 {\it J. Phys.} A {\bf 34} 8321

\bibitem{manko03}
Man'ko VI 2003 in {\it Theory of Nonclassical States of Light}, Dodonov V V and Man'ko V I eds. (Taylor \& Francis, London)

\bibitem{thom} 
Thom R 1972 {\it Stabilit\'e Structurelle et Morphog\'en\`ese}, (W.A. Benjamin, Reading, Mass.)

\bibitem{lambert04}
Lambert N, Emary C and Brandes T 2004 {\it Phys. rev. Lett.} {\bf 92} 073602

\bibitem{katzgraber}
Katzgraber H G {\it Phase Transitions: Proseminar in Theoretical Physics}, Institut f\"ur Theoretische Physik (ETH Z\"urich) SS07

\bibitem{SimInNat}
Casta\~nos O, Nahmad-Achar E, L\'opez-Pe\~na R and Hirsch J 2010 {\it AIP
  Conf. Proc.} {\bf 1323} 40

\bibitem{gelfand50}
Gelfand I M and Tsetlin M L 1950 {\it Dokl. Akad. Nauk SSSR} {\bf 71} 825 [english transl. in Gelfand I M 1988 {\it Collected Papers} Vol II (Springer Verlag, Berlin)]

\bibitem{Ramon}
For a detailed derivation cf. L\'opez-Pe\~na \etal ``Symmetry Adapted Coherent States for Three-Level Atoms Interacting with one-mode Radiation'' {\it Phys. Scr.} (to be published)

\bibitem{TriplePoint}
Nahmad-Achar E, Cordero S, L\'opez-Pe\~na R and Casta\~nos O 2014 {\it J. Phys. A: Math. Theor.} {\bf 47} 455301

\bibitem{erratum}
Casta\~nos O, Nahmad-Achar E, L\'opez-Pe\~na R and Hirsch J G 2011 {\it Phys. Rev.} A {\bf 84} 049901(E)

\end{thebibliography}
\end{document}